\documentclass[a4paper,12pt]{article}
\usepackage{graphicx}
\usepackage{hyperref}
\usepackage{epsfig}
\usepackage{amssymb}
\usepackage{setspace}

\usepackage{amsmath}
\usepackage{amssymb}
\usepackage{graphicx}
\usepackage{geometry}
\usepackage{subfigure}
\usepackage{url}
\usepackage{color}

\newcommand{\be}{\begin{equation}}
\newcommand{\ee}{\end{equation}}
\newcommand{\br}{\begin{eqnarray}}
\newcommand{\bea}{\begin{eqnarray}}
\newcommand{\eea}{\end{eqnarray}}
\newcommand{\er}{\end{eqnarray}}
\newcommand{\ba}{\begin{array}}
\newcommand{\ea}{\end{array}}
\newcommand{\bi}{\begin{itemize}}
\newcommand{\ei}{\end{itemize}}
\newcommand{\bn}{\begin{enumerate}}
\newcommand{\en}{\end{enumerate}}
\newcommand{\bc}{\begin{center}}
\newcommand{\ec}{\end{center}}

\newcommand{\ctll}{\cos\theta_{\ell\ell}}
\newcommand{\ctbb}{\cos\theta_{b\bar{b}}}
\newcommand{\detall}{\Delta\eta_{\ell}}
\newcommand{\detabb}{\Delta\eta_{b}}

 \def\({\left(}
 \def\){\right)}
 \def\[{\left[}
 \def\]{\right]}

\def\mHu{{m_{h_u}}}
 \def\mHd{{m_{h_d}}}
 \def\mHu2{{m_{h_u}^2}}
 \def\mHd2{{m_{h_d}^2}}

\def\gappeq{\mathrel{\rlap {\raise.5ex\hbox{$>$}}
{\lower.5ex\hbox{$\sim$}}}}

\def\lappeq{\mathrel{\rlap{\raise.5ex\hbox{$<$}}
{\lower.5ex\hbox{$\sim$}}}}

\begin{document}
\pagestyle{empty}
\begin{flushright}
{CERN-PH-TH/2014-037}
\end{flushright}
\vskip 2cm
\begin{center}
{\LARGE {\bf Enhancing the $t\bar tH$ signal through top-quark 
\\
\vspace{0.3cm} spin polarization 
effects at the LHC }}

\vspace*{1.5cm}
{\large
 {\bf Sanjoy~Biswas$^{a,b}$},
{\bf Rikkert~Frederix$^{c}$},
 {\bf Emidio~Gabrielli$^{{d,e,f}}$, and
} \\
\vspace{0.2cm}
  {\bf   Barbara~Mele$^{b}$}}
\vspace{0.3cm}

{\it
 (a)  Dipart. di Fisica, Universit\`a di Roma ``La Sapienza", \\ Piazzale Aldo Moro 2, I-00185 Rome, Italy}  \\[1mm]
 {\it
 (b) INFN, Sezione di Roma, Piazzale Aldo Moro 2, I-00185 Rome, Italy}  \\[1mm]
{\it
 (c) PH Department, TH Unit, CERN, CH-1211 Geneva 23, Switzerland}  \\[1mm]
 {\it
 (d) Department of Physics, University of Trieste, Strada Costiera 11, I-34151 Trieste, Italy}  \\[1mm]
{\it
 (e) NICPB, Ravala 10, Tallinn 10143, Estonia}  \\[1mm]
{\it
 (f) INFN, Sezione di Trieste, Via Valerio 2, I-34127 Trieste, Italy}

\vspace*{3cm}
{\bf ABSTRACT} \\
\end{center}
\vspace*{5mm}

\noindent
We compare  the impact of top-quark spin polarization effects 
in Higgs boson production  in association with  top-quark   pairs and in  corresponding  backgrounds at the LHC. Because of the spin-zero nature of the Higgs boson, one expects,  in the chiral limit for the top quarks,  
a substantial complementarity in 
$t\bar{t}$ spin correlations for a Higgs decaying into fermions/gauge-bosons  and $t\bar{t}$ spin correlations for the corresponding  irreducible $t\bar{t}f\bar f/VV$ backgrounds.
Although top mass effects in $t\bar t H$ production are in general dominant,  and seriously spoil the chiral-limit expectations, one can find observables that capture the $t \bar t$ angular spin correlations and can help in separating the signal from irreducible backgrounds. In particular, we show that,
for both  $H\to b\bar b$ and $H\to \gamma\gamma$,
 taking into account  $t\bar{t}$ spin correlations in  
$t\bar{t} H$ production  and  irreducible backgrounds could appreciably improve the LHC sensitivity to the $t\bar{t} H$ channel.

\vfill\eject

\setcounter{page}{1}
\pagestyle{plain}


\section{Introduction}
After the Higgs boson discovery in 2012 
\cite{Aad:2012tfa},\cite{Chatrchyan:2012ufa},
 with $m_H$ around 125 GeV,  the LHC experiments are due to test the new particle properties at the highest possible accuracy in the forthcoming years. The direct verification of the Yukawa sector describing the Higgs couplings to fermions in the standard model (SM) plays a major role in establishing  the actual nature of the observed new particle. The Higgs coupling $Y_t$ to the top quark, the heaviest fermion, is of particular phenomenological and theoretical interest, since on the one hand it rules the dominant one-loop production mechanism $gg\to H$, and, on the other hand, it governs the leading Higgs-mass corrections dependence on the theory cut-off energy scale, connected to the scalar-field naturalness problem in the SM.
 
Potential contributions from new virtual states circulating in the $gg\to H$, possibly induced by some unknown 
new physics,
  make the Higgs-top coupling  determination through $gg\to H$ production particularly model dependent.
A model-independent test of the $Y_t$
coupling relies instead on lower cross-section processes, notably  the Higgs boson production 
in association with a top-quark pair $pp\to t \bar{t} H$
\cite{Kunszt:1984ri}-\cite{Garzelli:2011vp}, where one can tag the actual presence of top quarks in the final state \cite{Maltoni:2002jr}-\cite{Belyaev:2002ua}.

The $t \bar{t} H$ production is a very challenging channel, traditionally thought to   require  the highest LHC collision energies and integrated luminosities in order to be discriminated from background.
The $t \bar{t} H$ cross section at  8-TeV $pp$ collisions is quite depleted by  the phase space of three heavy final particles, being just about
130 fb \cite{Dittmaier:2011ti}. Nevertheless, both the ATLAS and CMS experiments have been performing better than expected,  delivering a first set of quite constraining 
measurements of 
 $t \bar{t} H$ signal strengths for different Higgs decay channels \cite{ATLAS:2012cpa}-\cite{CMS:2013tfa}.
 Dedicated analysis have been done even in the rarest channels, like 
 $H\to \gamma\gamma$ and multileptons, where a more favorable signal-to-background ratio ($S/B$) compensates for the lack of statistics. The latter turn out to have a constraining power similar to the highest-statistics channels. 
 For instance, with the present data set of about 5 fb$^{-1}$ at 7 TeV plus about 20 fb$^{-1}$  at 8 TeV, a 95\% C.L. observed  upper limit
of 5.4 (5.3) times the SM cross section in the $H\to \gamma\gamma$ channel has been set by  CMS \cite{CMS:2013fda} (ATLAS \cite{TheATLAScollaboration:2013mia}), that is quite close to the corresponding  upper bound  for the $H\to bb$  channel based on the same data set  
by CMS \cite{CMS:2013sea}.
\\
 By combining the $H\to bb, \gamma\gamma, \tau\tau$ and multileptons analyses 
 \cite{CMS:2013fda}-\cite{CMS:2013tfa}, CMS  already delivered a first measurement of the $t \bar{t} H$ signal strength, based on the total present data set, as $\sigma/\sigma_{SM}\simeq 2.5^{+1.1}_{-1.0}$,
  assuming SM Higgs branching ratios  \cite{CMS:combi}.

The channel with highest statistic, arising from  $H\to b\bar{b}$,  suffers from large QCD backgrounds, mainly corresponding to the $t \bar{t} b \bar{b}$ and $t \bar{t}jj$ final states. The reconstruction of the $H\to b\bar{b}$ resonance is also
plagued by a combinatorial background arising from incorrect  $b$-jet assignment
(due either to extra $b$'s from $t$ and $\bar t$ decays or misidentified light jets).
The $t \bar{t}jj$ reducible background component amounts to more than 95\% of the total
\cite{Heinemeyer:2013tqa},
and can be normalized through control regions not contaminated by the signal. Requiring multiple $b$-jet tagging is also very effective in reducing it. On the other hand, the irreducible $t \bar{t} b \bar{b}$  component is hard to separate or fit through data-driven methods, being much smaller than, and kinematically very similar to, the dominant $t \bar{t}jj$. As a consequence, in order to separate the irreducible  background, it is crucial  to reach  the highest possible control of theory predictions for the
 $t \bar{t} b \bar{b}$ production. To this end, in  \cite{Heinemeyer:2013tqa} an up-to-date discussion on  the consistent interfacing next-to-leading-order (NLO) QCD perturbative predictions \cite{Bredenstein:2009aj}-\cite{Cascioli:2013era}\footnote{See also \cite{Bevilacqua:2010ve}-\cite{Hoeche:2014qda} for NLO QCD corrections to $t \bar{t} jj$.} with parton showers at 14 TeV is presented. After applying, a particular set of cuts optimizing the signal to background ratio,  present analysis foresee  $S/B\lappeq 1/20$. Developing  strategies aimed to better discriminate the $t \bar{t} H$ signal distributions versus the irreducible $t \bar{t} b \bar{b}$
 background is hence crucial, particularly in view of the much higher statistics  that will be accumulated in  forthcoming years at 14 TeV.

 The scope of the present study is to explore the potential of the spin-correlation properties in the associated Higgs top-pair production at the LHC as a possible tool to improve the separation of  the signal from the $t\bar{t}H$ irreducible backgrounds. 
The  $t\bar{t}H$ spin properties have recently been considered in the literature as a mean to characterize a SM signal versus possible non-SM effects. In particular, spin correlations  could help  
in disentangling the SM scalar component from a  pseudoscalar contribution in the 
top-Higgs coupling \cite{Ellis:2013yxa}. In \cite{Artoisenet:2012st} it was   emphasized 
  that the relative impact of spin correlations on the leading-order (LO) $t\bar{t}H$ lepton kinematical distributions is  much more dramatic than the one of the corresponding QCD NLO corrections~\cite{Frederix:2011zi}.

The top quark is unique among all quarks, 
its life time being shorter than the characteristic hadronization time scale.
Top quarks are then expected to decay before their original spin is affected by strong interactions, so ensuring 
 that spin polarization
at production level is fully transferred to the top decay products. 
Hence, by reconstructing the individual top systems (which can actually be done even in presence
of two neutrinos in the final state \cite{ATLAS:2012ao}), 
 the  top-quark spin properties can be accessed by measuring  angular 
distributions of the final  decay products in $t\to W+b\to \ell \nu(du)+b $ \cite{Beneke:2000hk}. Among the top decay products, the charged lepton (or $d$ quark) has the maximal spacial correlation with the original top-quark spin axis 
\cite{Jezabek:1994qs}-\cite{Brandenburg:2002xr}.

Although in the SM top quark and antiquark pairs  are mostly unpolarized 
in $\bar t t$ production 
 at hadron colliders, their spins   are strongly correlated
  \cite{Bernreuther:2001rq}-\cite{Bernreuther:2010ny}, as  confirmed by  present experimental studies  at the Tevatron \cite{Aaltonen:2010nz},\cite{Abazov:2011gi} and the LHC 
   \cite{ATLAS:2012ao},\cite{Chatrchyan:2013wua}.

In a naive picture, in the chiral limit of vanishing top-quark mass (or for 
very high invariant masses of the  $t\bar{t}$ system, $m_{tt}\gg m_t$)
the top quark and antiquark spins are highly correlated  and parallel to each other along the $t\bar{t}$ production axis. Top pairs are hence
produced in the LR + RL helicity configurations, where L(R) stands for the left(right)-handed helicity polarization.
In the same kinematical limits,
when the $t\bar{t}$ is produced in association with a Higgs boson,
the top quark  and antiquark helicities are also correlated, but in a 
complementary way with respect to the previous case. 
Indeed the Higgs-boson emission from the 
top-quark final states via Yukawa interactions   
induces a chirality flip in the top-quark polarization.
Then, in contrast with $t\bar{t}$ production, for large $m_{tt}$ invariant masses the dominant $t\bar{t}$ helicity configurations in the $t\bar{t}H$ final state will be  LL+RR, while the LR+RL configuration is expected to be suppressed by terms of order 
${\cal O}(m_t^2/m_{tt}^2)$. 

We can now extend 
 the same  top-quark chiral limit to the irreducible backgrounds for $t\bar{t}H$. For $H\to\gamma\gamma$, the irreducible background arises from 
direct $t\bar{t}\gamma\gamma$ production. In this case, the emission of photons
from the quark lines in the $gg\to t\bar{t}$ and $ q\bar{q}\to t\bar{t}$ amplitudes is not expected to affect  the basic LR+RL helicity correlation of the plain top-quark pair
production. One then should have some complementarity  in the spin-correlation  properties of 
 the $t\bar{t}H(\to\gamma\gamma)$ signal and the ones of the $t\bar{t}\gamma\gamma$ irreducible background.
 As for  the decay channel $H\to b \bar{b}$, the $t\bar{t}H(\to bb)$ signal presents of course the same spin correlations as the ones in the   $t\bar{t}H(\to\gamma\gamma)$ channel. On the other hand,  the analysis of the $t\bar{t}b\bar{b}$ irreducible background is  less straightforward than in $t\bar{t}\gamma\gamma$ even in the top chiral limit,
 since many different topologies (presenting in general different $t\bar{t}$ spin correlations) contribute to the  $t\bar{t}H(\to bb)$ amplitude. Nevertheless, one expects that the action of the Yukawa coupling in the signal channel should leave some imprint  diversifying  the 
 signal polarization features of $t\bar{t}H(\to bb)$ with respect to the $t\bar{t}b\bar{b}$ irreducible background also in this case.
 
Just as it actually happens for the $t\bar{t}$ production \cite{Mahlon:2010gw}, when dropping the unrealistic (for LHC energies) top chiral-limit assumption, predicting polarization properties of the $t\bar{t}H$ signal and corresponding backgrounds gets  much harder. Top mass effects are indeed dominant in the bulk production of top pair systems. In $t\bar{t}$ production, the structure of spin correlations changes significantly over the top production phase space, with modulations  (arising from the interference of different helicity states) that widely vary from threshold production to boosted-top regime  \cite{Baumgart:2012ay}. As a consequence, the top-antitop 
spin-correlation properties in $t\bar{t}H$ and corresponding irreducible backgrounds are in general not simple to predict on the basis of the previous naive arguments. 

In this paper, we  
study LO distributions of top decay products in  $t\bar{t}H$ versus corresponding irreducible backgrounds, by keeping the correct correlation effects between top polarization in production  and decay. We 
 try to identify polarization observables that are particularly sensitive to separate the signal from irreducible backgrounds.  The dependence on the reference-frame choice is also discussed.

We focus on the two  
channels corresponding to the $H\to\gamma\gamma$ and  
 $H\to bb$  decays
 \bea
pp&\to& t\;\bar t \;H\;(\to \!\gamma\gamma) \\
\label{togamgam}
pp &\to& t\;\bar t \;H\;(\to bb) \, ,
 \label{tobb}
 \eea 
where the corresponding irreducible backgrounds $t\bar{t}\gamma\gamma$ and $t\bar{t}b\bar{b}$ are expected to play a major role with respect to reducible ones.

Note that, regarding the  $t\bar{t}\gamma\gamma$  background, 
there are extra contributions to the observed final state $\ell\nu b\; \ell \nu b \;\gamma\gamma$
due to  photon brehmsstralung from charged top decay products \cite{Buttar:2006zd}. The latter can fake  
the irreducible $t\bar{t}\gamma\gamma$ background, and affect the reconstruction of the spin properties  of   $t\bar{t}$ pairs. Proper kinematical cuts on the decay products of the top quarks can help to reduce these contributions, 
although eventual $t\bar{t}$ spin correlations 
are in general quite affected by  kinematical cuts\footnote{ Similarly, $b$ pairs  generated via  gluon radiation   from $b$ arising in top decays
give  extra contributions to the  $\ell\nu b\; \ell \nu b \;b\bar{b}$  background from $t\bar{t}b\bar{b}$.
The latter mildly contribute due to the corresponding low $b\bar b$ invariant mass.}.
 NLO QCD corrections to $t\bar{t}\gamma\gamma$ could be further incorporated through the MadGraph5\_aMC@NLO framework \cite{MadaMC}.

In the Section 2, we  start with a more quantitative analysis of the polarization effects in the $t\bar{t}H$ production. We then proceed to a detailed analysis of the signal and  corresponding irreducible backgrounds 
for  $H\to \gamma \gamma$ and $H\to b\bar{b}$,
 in Section 3 and 4, respectively. We then draw our conclusions in the Section 5.
\section{Spin correlations in $t\bar{t}H$ production }

\begin{figure}[hp]
\begin{center}
\includegraphics[width=0.49\textwidth]{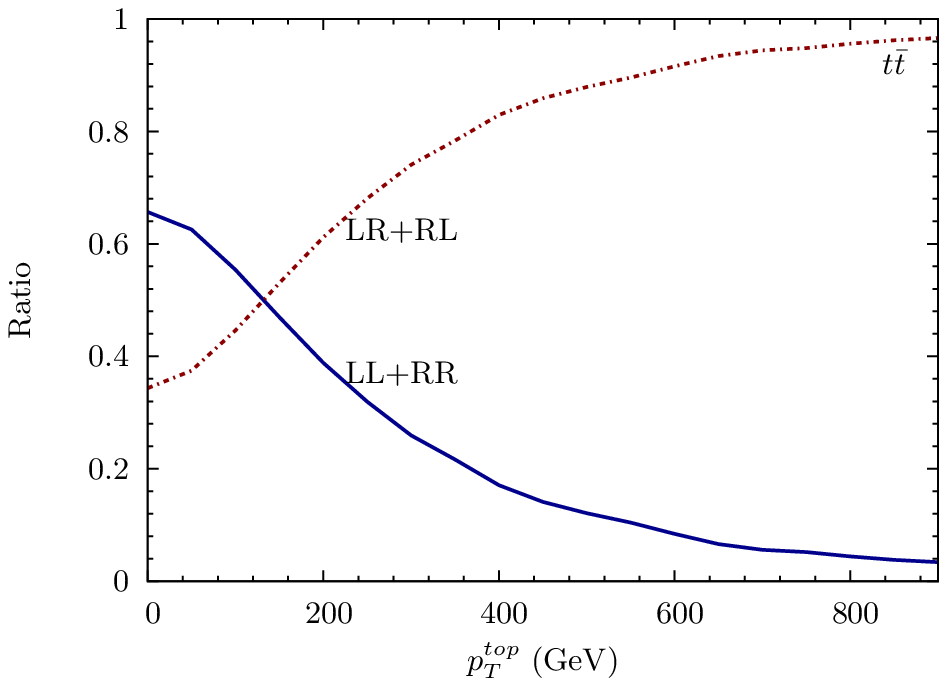}\hskip 9pt
\includegraphics[width=0.49\textwidth]{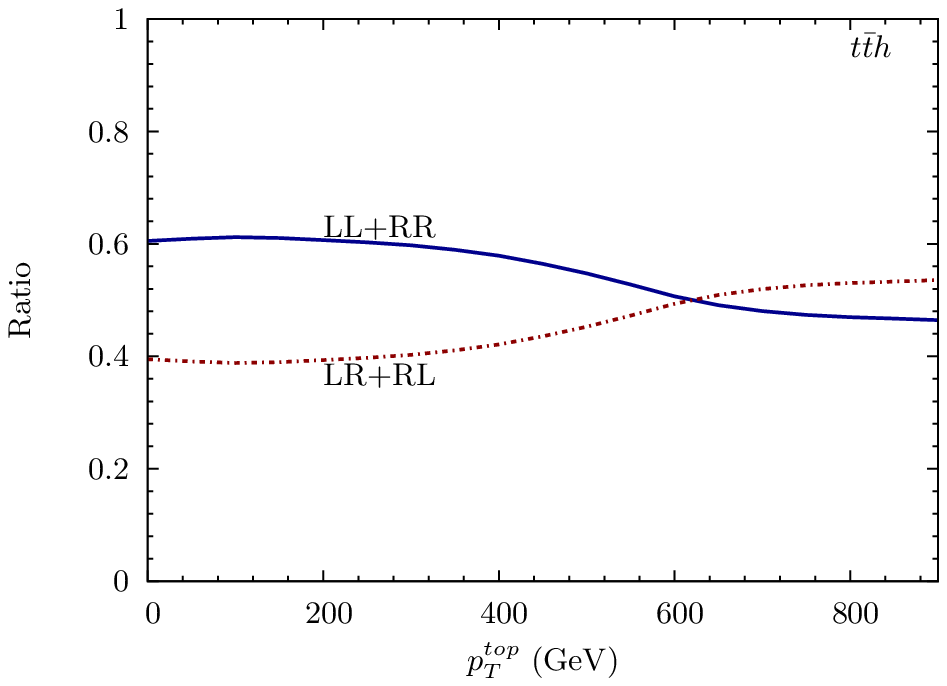}
\caption{Integrated $p_T$ distributions for the   like-helicity top pairs ($t_L\bar t_L+t_R\bar t_R$) and unlike-helicity top pairs ($t_L\bar t_R+t_R\bar t_L$)
in  unpolarised {\bf $t\bar{t}$} (left plot) and 
{\bf $t\bar{t}h$} (right plot) samples, versus the  hardest-top $p_T$ cut at c.m. energy of 14 TeV, in the Lab frame.}
\label{Fig1}
\end{center}
\end{figure}

Before entering the details of the signal and background analysis for the $t\bar{t}H$
production, we show in Fig.~\ref{Fig1}, 
the integrated top $p_T$ distributions projected on the 
$t\,\bar t$ LL+RR and LR+RL helicity configurations (normalized to the total cross section),  
for the $t\bar{t}$ (left plot) and $t\bar{t}H$ (right plot) productions, in the Lab frame. Here, 
  $p_T^{top}$ is the minimum  transverse momentum of the hardest top. 
We can see that helicity correlations as a function of $p_T^{top}$ (which is directly connected to the   invariant mass $m_{tt}$ of the  $t\bar{t}$ system  adopted before)
for $t\bar{t}$  and $t\bar{t}H$ productions 
are not complementary as in the top chiral limit, but present a quite different behavior.
In $t\bar{t}$ production, one indeed observes the chiral limit emerging at large $p_T^{top}$, with the LR+RL component saturating the production rate.
On the other hand,  in the $t\bar{t}H$ production one does not have a complementary LL+RR dominance
at high $p_T^{top}$ as expected from   the 
Higgs-sstralung chirality flip discussed in Section 1. Top mass effects 
and the presence in the final state of a further massive (Higgs) particle makes the 
LL+RR and LR+RL  helicity
components in $t\bar{t}H$ less unbalanced in all the statistically relevant top 
transverse-momentum range.
One expects that the actual chiral limit is reached in $t\bar{t}H$ production for much higher $p_T^{top}$ values than the  ones that can be experimentally covered with the actual LHC energy and luminosity. While in $t\bar{t}$ production the fraction of
the chiral-limit LR+RL configuration is above 80\% for $p_T^{top}\gappeq 400$ GeV,
in $t\bar{t}H$ it moderately gets the upperhand for $p_T^{top}\gappeq 600$ GeV, effectively dampened by Higgs radiation.

\begin{figure}[t]
\includegraphics[width=0.49\textwidth]{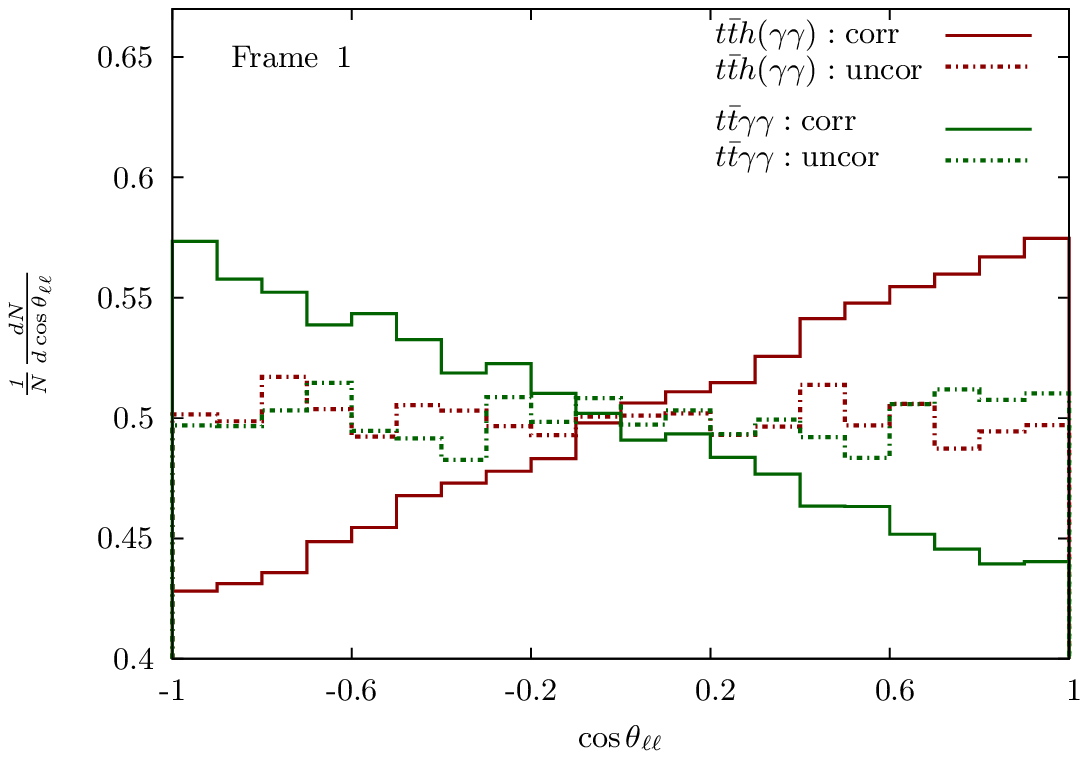}\hskip 9pt
\includegraphics[width=0.49\textwidth]{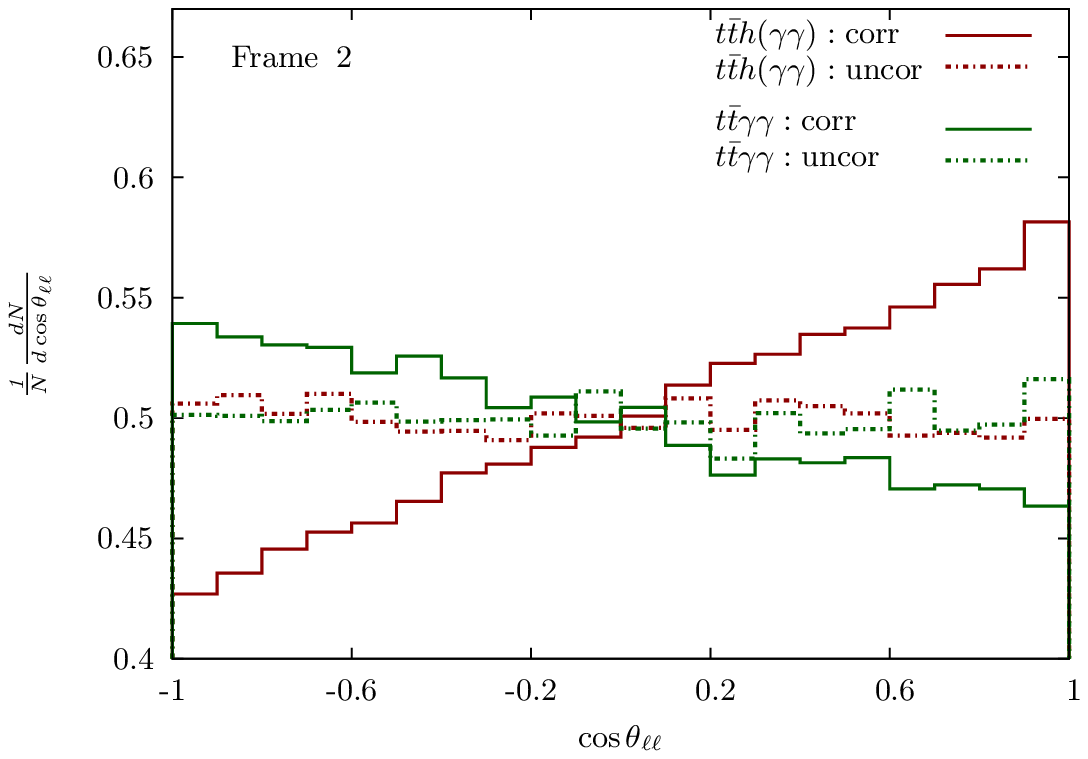}
\caption{ The 
$\ctll$ 
distribution for the signal $t\bar{t}H (H\to \gamma\gamma)$ (red) and  $t\bar{t}\gamma\gamma$ background (green), with (solid) and without (dashed) spin information,  in \text{Frame 1} (left) and \text{Frame 2} (right). The cuts 
$p^{\gamma_{1,2}}_T> 20$ GeV, $|\eta_{\gamma_{1,2}}|<2.5$ and $\Delta R_{\gamma_1\gamma_2} > 0.4$ have been imposed 
on  photons, in addition to the invariant mass cut $123{\rm ~GeV}< m_{\gamma\gamma}< 129 {\rm ~GeV}$.}
\label{Fig2}
\end{figure}

We then compare the fraction of LL+RR and LR+RL  helicity configurations that contribute to the total $t \bar t H$ cross section, with the corresponding fractions in the
irreducible backgrounds corresponding to the $H\to\gamma\gamma$ and  
 $H\to bb$ channels.
 We find that about 61\% of the total $t \bar t H$ cross section corresponds to
 the LL+RR combination, with a remaining 39\% for  LR+RL.
  Regarding the  $t\bar{t}\gamma\gamma$  background (with  basic cuts defined in the  Fig.~\ref{Fig2} caption later on), we have 28\% of the total  cross section with  LL+RR combination, and  a remaining 72\% for LR+RL.
   As for  the  $t\bar{t}b \bar b$  background (with  basic cuts defined in the  Fig.~\ref{Fig7} caption later on), we have an almost equal fraction  of LL+RR and LR+RL configurations.
   One can see that in the $t\bar{t}\gamma\gamma$  background one indeed finds an opposite trend in the helicity configurations with respect to the $H\to\gamma\gamma$ signal. In the $t\bar{t}b \bar b$  background the effect is washed out in the integrated cross section.
 
We now proceed to scrutinize the effects of $t\bar{t}H$ top spin properties through
the study of angular variables involving  decay products of the top pair into two semileptonic final states $t \to b\ell \nu$, $\bar t \to \bar b\ell \nu$. We will focus in particular
on charged-lepton pair and $b$-jet pair variables, where the $b$ jets originate from top and antitop decays, applying strategies previously explored in $t\bar{t}$ production.
\\
Spin correlations  in  hadronic $t\bar{t}$ production can be measured by studying 
 angular distributions of  top decay products in specific frames and 
coordinate basis \cite{Bernreuther:2001rq}-\cite{Bernreuther:2010ny}. 
The following observables and corresponding distributions have 
been extensively used\footnote{For a recent discussion, see\cite{Baumgart:2012ay}.}:

\begin{itemize}
 \item Double polar decay distributions ($\frac{d^2\sigma}{d\cos\theta \,d\cos\bar \theta}$),  where $\theta$ ($\bar \theta$) is the polar 
  angle of any  of the top (antitop) decay products.

\item Distributions of three-dimensional opening angles ($\varphi$) between one top  decay products and  one  anti-top decay products ($\frac{d\sigma}{d\cos\varphi}$).

\item More compact one-dimensional observables given by combinations of azimuthal angles $\phi \pm \bar\phi$ of  two top and antitop 
decay products  ($\frac{d\sigma}{d(\phi \pm \bar\phi)}$).

\end{itemize}

In this work we focus on the second kind of observables  as they capture most of the 
$t\bar{t}$-system spin-correlation effects at moderate values of the  top-quark  $p_T$
\cite{Baumgart:2012ay}. 
Apart from the laboratory frame, we will consider  two further reference frames. The latter, although unusual,  have been introduced in 
\cite{Bernreuther:2004jv}, being particularly  sensitivite to the 
$t\bar t$ spin correlations compared to the lab frame.

We define   
the angle between the direction of flight of $\ell^+$ ($b$) in the $t$ rest system and 
$\ell^-$ ($\bar b$) in the $\bar t$ rest system \cite{Bernreuther:2004jv}. 
Two different rest systems are involved in the above angular-variable definition, and 
to avoid  ambiguities one has to specify the common initial frame where  Lorentz boosts are applied to separately bring the $t$ and $\bar t$ at rest.

The two aforementioned frames are defined as follows. The   $t$  and $\bar t$ rest systems are obtained by two corresponding rotation-free Lorentz boosts :
\begin{itemize}
\item 
(Frame-1) with respect to
 the $t\bar{t}$-pair c.m. frame,

\item
(Frame-2) with respect to the laboratory  frame.

\end{itemize}

In this paper we analyze the $t\bar{t}$ 
spin-correlations in both
 signal and background for two relevant Higgs decay channels $H\to \gamma\gamma$ and $H\to b\bar{b}$. 
For both  signal 
and background the $t \to b\ell \nu$ decay has been performed in MadGraph5 \cite{Alwall:2011uj} 
by retaining the full spin information, while  
in the uncorrelated case, the decay has been included by interfacing MadGraph5 with PYTHIA \cite{Sjostrand:2006za} at the $t \bar t H$ level, so neglecting the spin polarization effects. In our analysis, 
we do not include shower nor hadronization effects. 
In the following,  we will show the results 
for  angular and rapidity distributions
for  the 
decay products of the top and  antitop, in both the signal $t\bar{t}H$, with 
$H\to \gamma \gamma$ and $H\to b\bar{b}$, and corresponding irreducible backgrounds $t\bar{t}\gamma\gamma$ and $t\bar{t}b\bar{b}$, respectively.

The variables that we found particularly sensitive to spin correlations are
$\ctll$  and $\detall$ (with  
$\detall\equiv |\eta_{\ell^+}-\eta_{\ell^-}|$), 
where $\theta_{\ell\ell}$ is the three-dimensional polar angle between the $\ell^+$ and  $\ell^-$ directions of flight, and  $\eta_{\ell}$ is the pseudo rapidity of individual leptons  in a specific frame.
Analogously, we adopt the $\ctbb$  and $\detabb$  variables   
(with $\detabb\equiv |\eta_{b}-\eta_{\bar{b}}|$) involving the two $b$ quarks from $t$ and $\bar t$ decays.
This set of variables has been pinpointed after a careful scrutiny including different 
proposals  
adopted in $t \bar t$ production studies \cite{Mahlon:2010gw},\cite{Baumgart:2012ay}.

In all the following plots, the signal (background) distributions will be shown via red (green) lines, while spin-correlated (-uncorrelated) cases will be reported through solid (dashed) lines. All distributions are normalized to 1.  In the following,  in each case we will show  distributions  that we found to be particularly sensitive to spin effects.
\section{The $t\bar t \gamma \gamma$ channel}
\begin{figure}[t]
\begin{center}
 \includegraphics[width=0.49\textwidth]{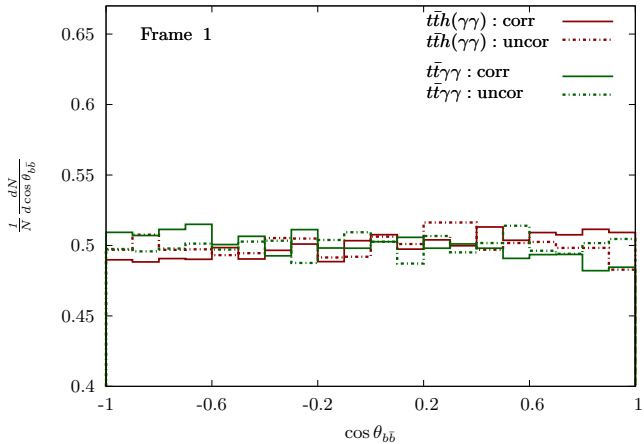}
\caption{ The 
$\ctbb$  
distribution for the signal $t\bar{t}H (H\to \gamma\gamma)$ (red) and  $t\bar{t}\gamma\gamma$ background (green), with (solid) and without (dashed) spin information,  in \text{Frame 1}. Same cuts as in Fig.~2 have been imposed.}
\label{Fig3}
\end{center}
\end{figure}

\begin{figure}[ht]
\includegraphics[width=0.49\textwidth]{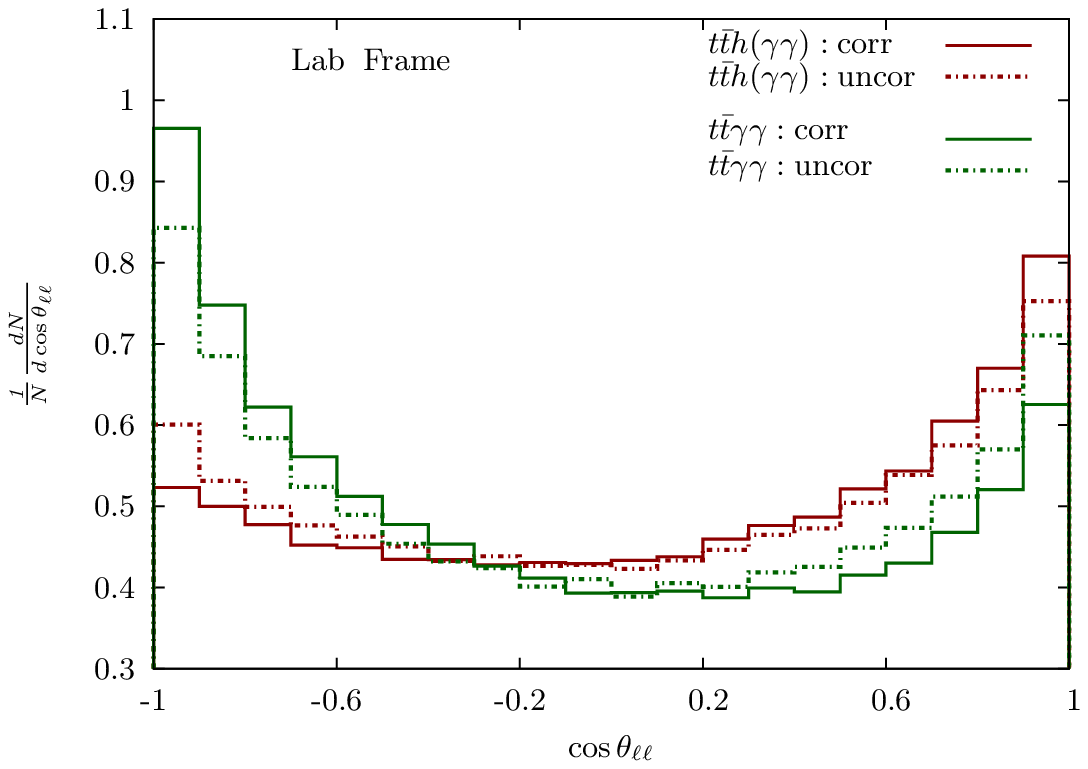}\hskip 9pt
\includegraphics[width=0.49\textwidth]{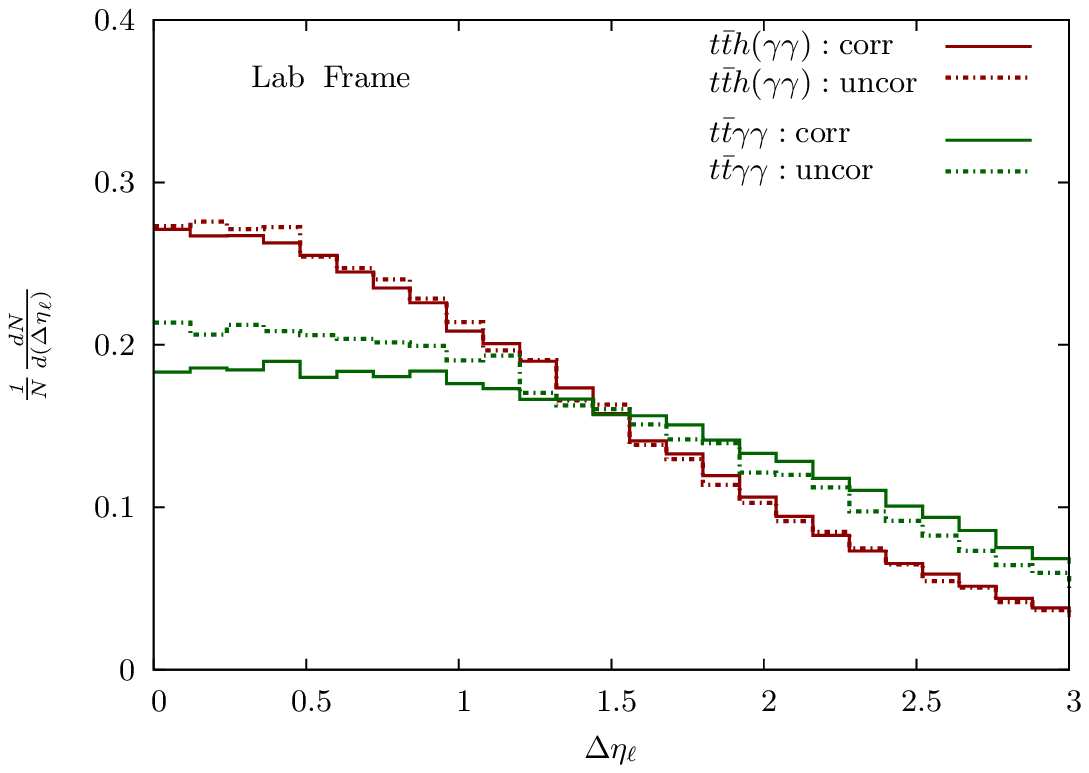}
\vskip 5 pt
\includegraphics[width=0.49\textwidth]{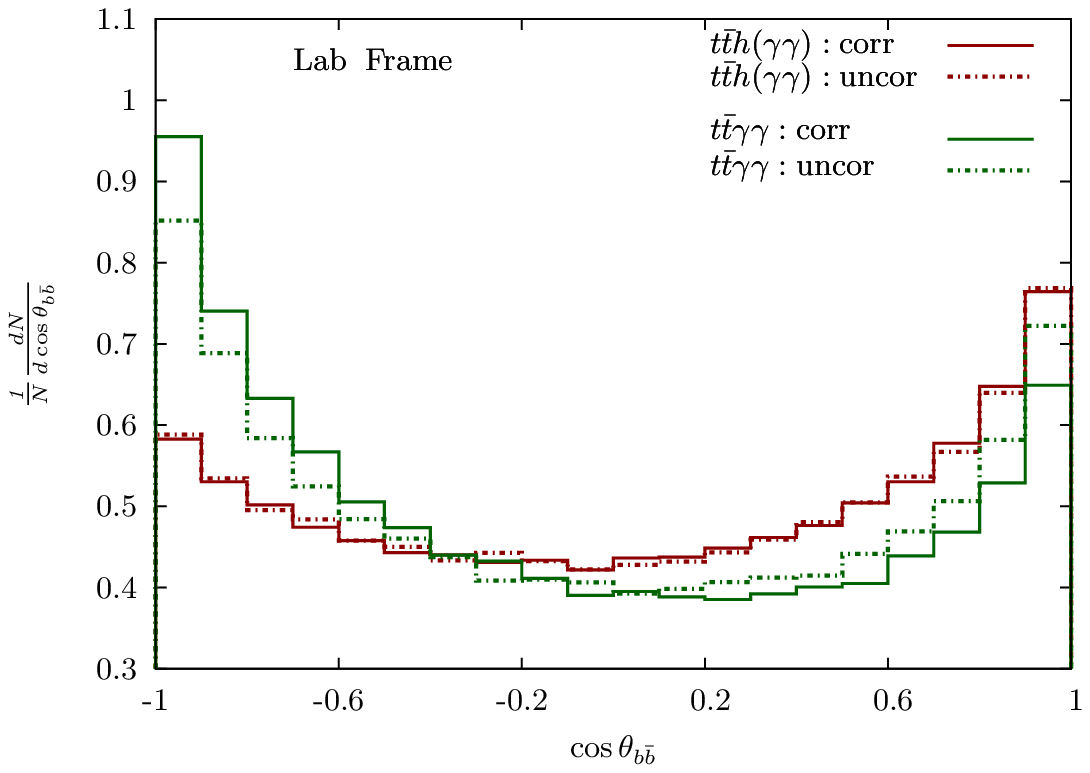}\hskip 9pt
\includegraphics[width=0.49\textwidth]{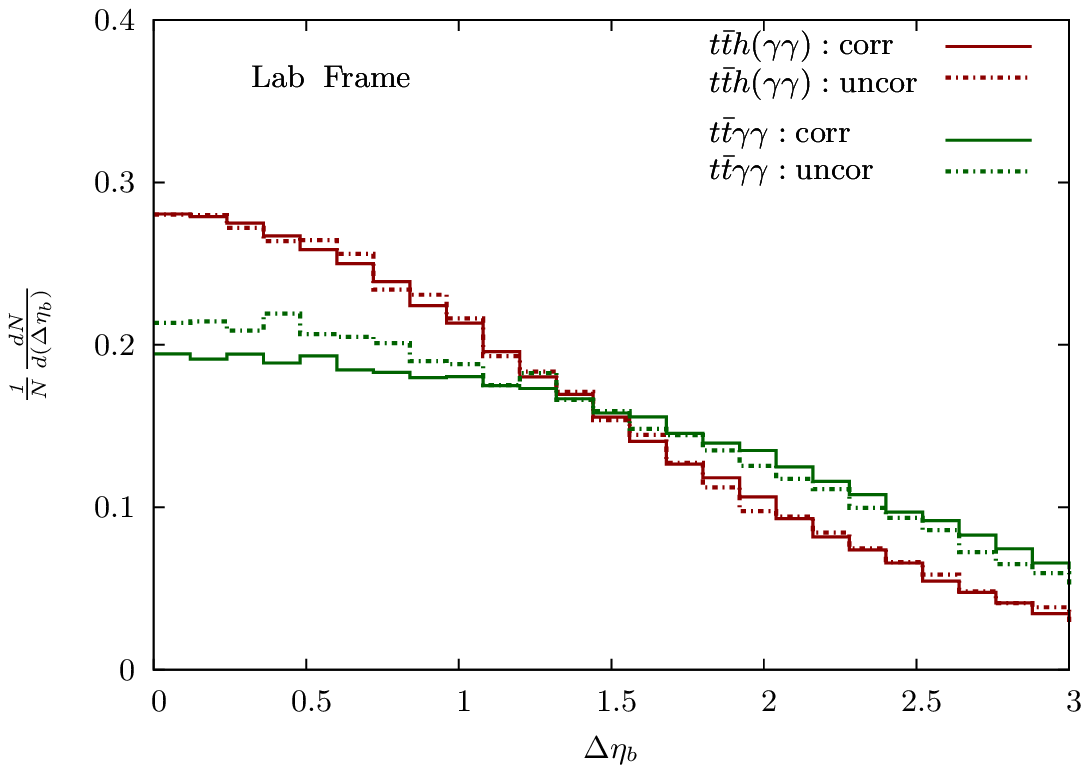}
\caption{ The 
$\ctll$ (top left), $\detall$ (top right),  $\ctbb$ (bottom left),  and $\detabb$ (bottom right) 
distributions for the signal $t\bar{t}H (H\to \gamma\gamma)$ (red) and  $t\bar{t}\gamma\gamma$ background (green), with (solid) and without (dashed) spin information, in the \text{Lab frame}. Same cuts as in Fig.~2 have been imposed.}
\label{Fig4}
\end{figure}

\begin{figure}[t]
\includegraphics[width=0.49\textwidth]{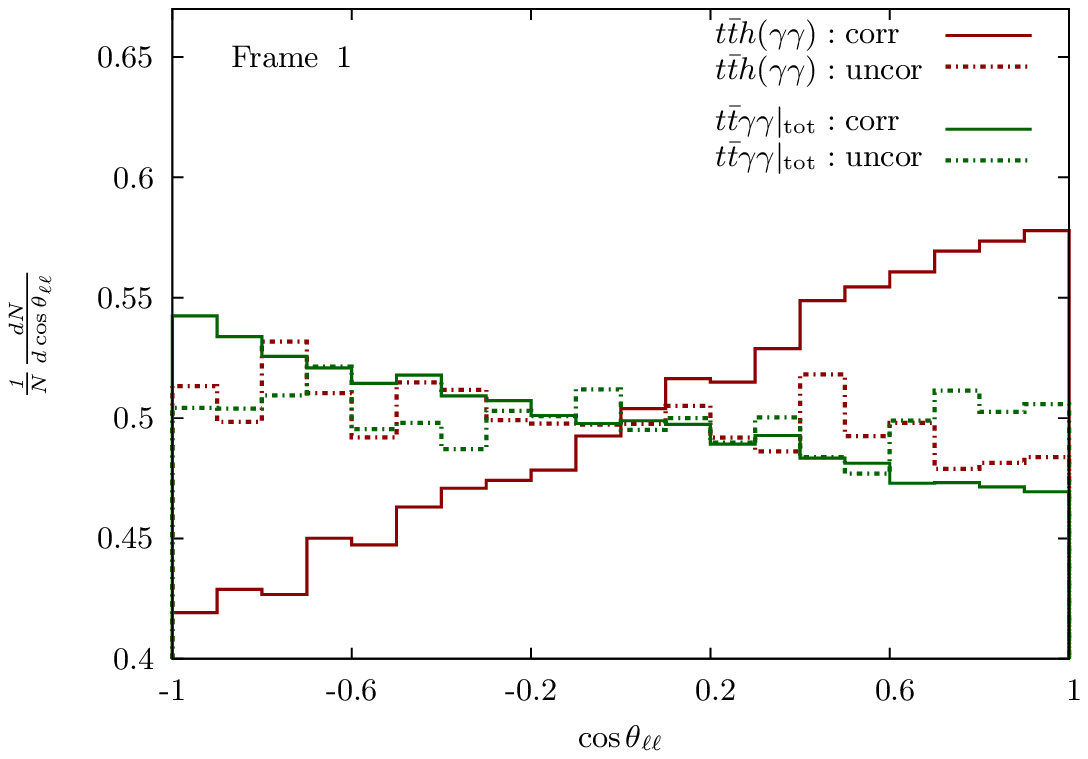}\hskip 9pt
\includegraphics[width=0.49\textwidth]{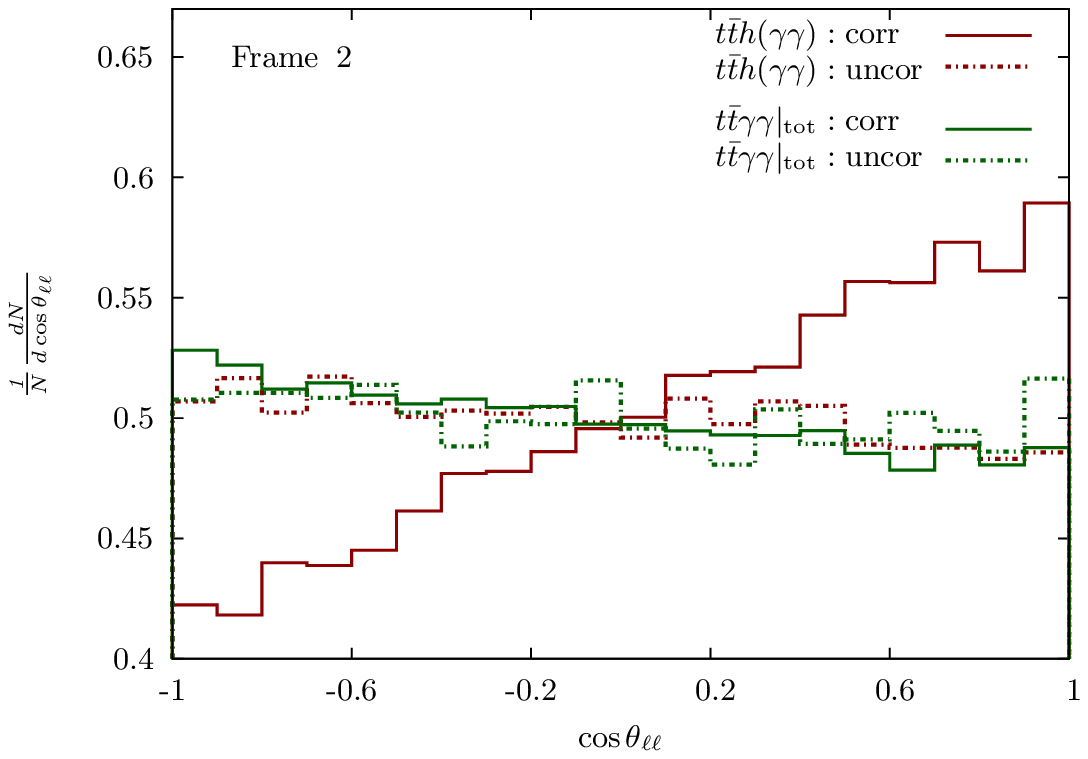}
\caption{ The 
$\ctll$ 
distribution for the signal $t\bar{t}H (H\to \gamma\gamma)$ (red) and  
full $\gamma\gamma$ background (including radiation from $t$,$\bar t$ decay products, as defined in the text) (green), with (solid) and without (dashed) spin information,  in \text{Frame 1} (left) and \text{Frame 2} (right).  The set of cuts listed in 
Eq.(\ref{Cuts-1}-\ref{Cuts-4}) have been applied to both  signal and background.}
\label{Fig5}
\end{figure}

\begin{figure}[t]
\includegraphics[width=0.49\textwidth]{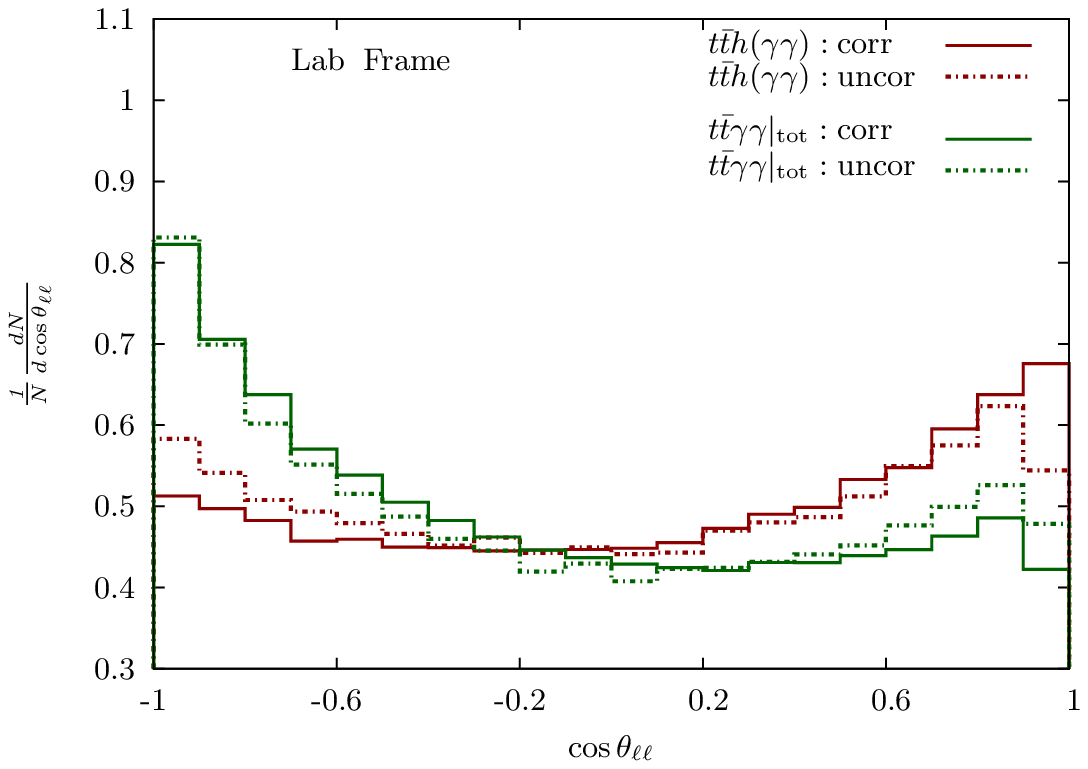}\hskip 9pt
\includegraphics[width=0.49\textwidth]{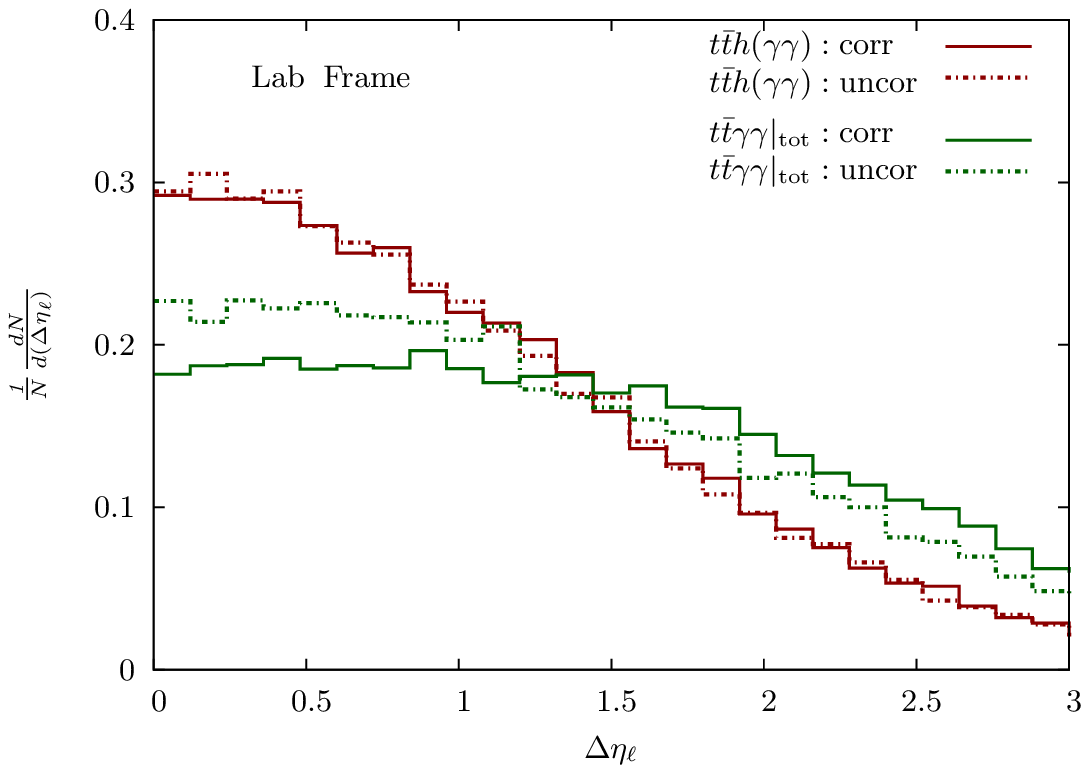}
\vskip 5pt
\includegraphics[width=0.49\textwidth]{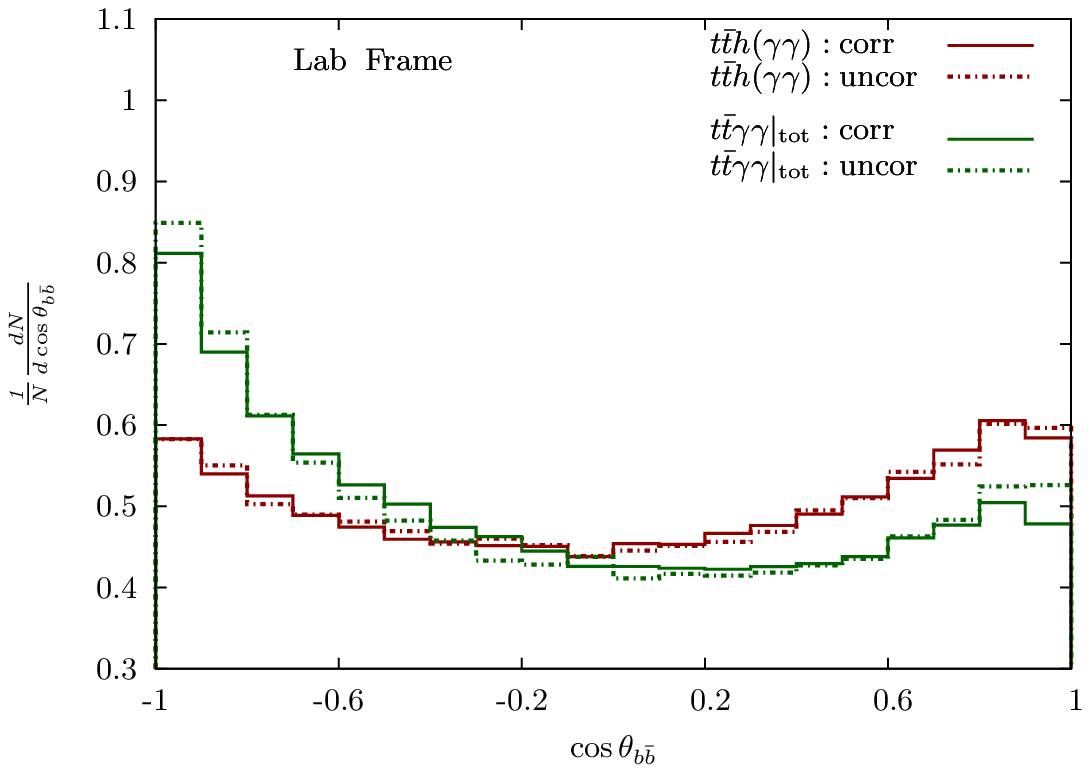}\hskip 9pt
\includegraphics[width=0.49\textwidth]{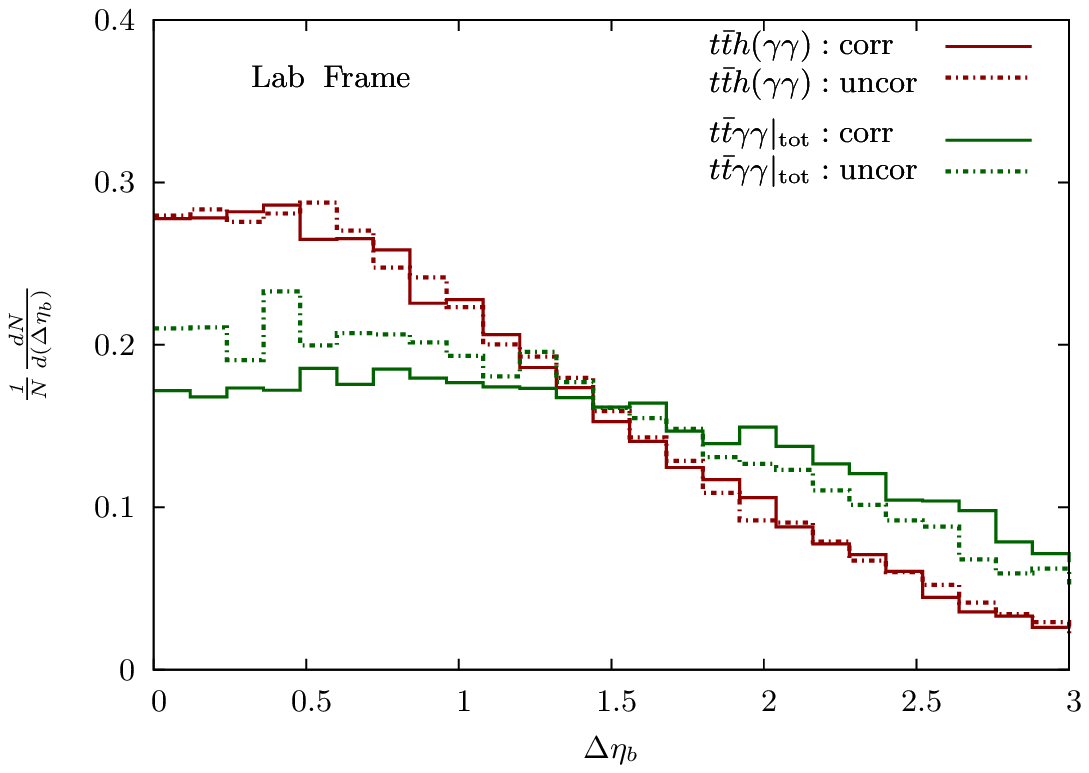}
\caption{ The 
$\ctll$ (top left), $\detall$ (top right),  $\ctbb$ (bottom left),  and $\detabb$ (bottom right) 
distributions for the signal $t\bar{t}H (H\to \gamma\gamma)$ (red) and  full $\gamma\gamma$ background (including radiation from $t$$\bar t$ decay products, as defined in the text) (green), with (solid) and without (dashed) spin information, in the \text{Lab frame}. Same cuts as in Fig.~5 have been imposed.}
\label{Fig6}
\end{figure}

Results for  $t\bar t \gamma \gamma$  are shown 
in  Figs.~\ref{Fig2}-\ref{Fig6}.
Distributions in 
Figs.~\ref{Fig2}-\ref{Fig4} include in the irreducible background  
only the photons emitted by
 initial and final charged states  in the $gg,q\bar q \to t\bar t$ partonic precesses.
In Figs.~\ref{Fig5}-\ref{Fig6}, 
the contributions from  photons emitted by the $t$ and $\bar t$ charged decay products 
are also included. 

In Fig.~\ref{Fig2},  we show the $\ctll$ distributions for $t\bar t \gamma \gamma$ signal 
 and irreducible background  with and without 
correlations in the  $t,\bar t$ spin polarizations. 
In the left and right plots we report 
the results in the  \text{Frame-1} 
and \text{Frame-2}, respectively, as defined above. Here, we impose just  
the following kinematical cuts
on the  photons   transverse 
momenta, $p^{\gamma_{1,2}}_T > 20$ GeV, pseudorapidities, 
$|\eta_{\gamma_{1,2}}|<2.5$, and isolation, $\Delta R_{\gamma_1 \gamma_2} > 0.4$,
in addition to a diphoton invariant mass cut $123{\rm ~GeV}< m_{\gamma\gamma}< 129 {\rm ~GeV}$, where $\Delta R_{ij}$ is as usual 
$\Delta R_{ij}=\sqrt{\eta_{ij}^2+\phi_{ij}^2}$, with $\eta_{ij}(\phi_{ij})$ 
the rapidity (azimuthal) separation.

In the  uncorrelated
analysis, the angular distributions for the signal and  background are
both flat in $\ctll$, in both reference frames. On the contrary, 
when the spin information is taken into account, the signal and background
distributions  are different and almost
complementary. In particular, the  signal (background) 
 distributions 
is  monotonically increasing (decreasing) as a function of $\ctll$.   This is  a  consequence of 
the aforementioned complementarity in the $t\bar{t}$ helicity correlations
of the signal and irreducible background for the $H\to \gamma\gamma$ channel,
that we have previously discussed.
Although the correlation effect is remarkable both in Frame-1 and Frame-2, the separation between the correlated $\ctll$ distributions 
for signal and background is  more enhanced in 
 Frame-1, where one gets an improvement in  $S/B$ (computed by integrating angular distributions over the range $0<\ctll<1$) 
 of about 17\%, compared to the uncorrelated case.
 
 In Fig.~\ref{Fig3} , we show the corresponding distributions in the  $\ctbb$ variable in Frame-1 (analogous results hold in Frame-2). The distributions are all approximately flat, and no significant effect is found in this case.

In Fig.~\ref{Fig4}, for comparison, we consider various  distributions  in the laboratory frame,
where the variables studied are more straightforward to reconstruct experimentally.
We analyze  the correlated and uncorrelated distributions in $\ctll$ and $\detall$ (top), and
$\ctbb$  and $\detabb$ (bottom)
(with $\detall\equiv |\eta_{\ell^+}-\eta_{\ell^-}|$), $\detabb\equiv |\eta_{b}-\eta_{\bar{b}}|$).
In all plots in the Lab Frame  the inclusion of spin correlations  increases the difference in  distribution 
shapes between signal and background, although the relative effect is quite smaller than in Frame 1 and Frame 2 for leptonic distributions. The $\ctbb$ distribution in the Lab Frame is instead much more effective in separating signal and background with respect to 
 Frame 1 (where it is almost flat in all cases cf. Fig.~\ref{Fig3}). Distributions in rapidity separations turns out to be more selective in the 
 $\detall<1.5$ range.

Now, we consider the effects induced including the contributions of
 photon emissions from the $t$ and $\bar t$ charged decay products 
 in the irreducible $\gamma\gamma$ continuum.
In order to do so, additional kinematic cuts are required 
for photon and $b$-jet isolation
\bea
 &p^{b,\gamma_{1,2}}_T> 20 {\rm ~GeV} {\rm ~and~} p^{\ell}_T> 10 {\rm ~GeV}&
\label{Cuts-1}\\
\nonumber\\ \ 
& |\eta^{b}|< 4.7, |\eta^{\ell}| < 2.7 {\rm ~and~} |\eta^{\gamma}|< 2.5& \label{Cuts-2}\\
\nonumber\\
&\Delta R (bb, \ell\ell, \gamma\gamma, b\, \ell, b\gamma, \ell \gamma) > 0.4 &
\label{Cuts-3} \\\nonumber\\ 
&  123 {\rm ~GeV} <m_{\gamma\gamma}< 129 {\rm ~GeV}\, .& \label{Cuts-4}
\eea
In Fig.~\ref{Fig5} we show the effects of including both photon radiation from 
$t$,$\bar t$ decay products and the new set of kinematical cuts defined in 
Eq.(\ref{Cuts-1}-\ref{Cuts-4}) on signal and background. We show   
the $\ctll$ distributions in Frame-1 (left) and Frame-2 (right), to be compared
to the ones not including extra background radiation in Fig.~\ref{Fig2}.
One can see that the $\ctll$ distributions for signal are basically unaffected by the new selection  cuts, while in  the background the extra photon radiation tends to reduce the gap between 
the correlated and uncorrelated $\ctll$ distributions.  In Frame-1 one gets an improvement
by 14\% in $S/B$.

In the Lab frame,  Fig.~\ref{Fig6} shows the  $\ctll$ (top left), $\detall$ (top right),
$\ctbb$ (bottom left)  
and $\detabb$ (bottom right) distributions  including extra photon radiation and new selection cuts.
One can see that the effects
of  photon emission from the top decay products do not dramatically 
affect the previous results where these contributions where ignored (cf. Fig.~\ref{Fig4}).
Differences are found mainly for low separations of lepton and $b$ pairs
(that is for $\ctll,\ctbb\sim 1$ and $\detall,\detabb<1$), where the new set of cuts is more effective.
In conclusion,  we find that the analysis of the $\ctll$ distributions for the channel $t\bar{t}H (H\to \gamma\gamma)$ and its irreducible background in  a study that  correctly takes into account 
 spin-correlation effects could significantly help in enhancing the signal-to-background ratio with respect to the uncorrelated analysis. The use of dedicated reference frames
 for reconstruction, like 
 Frame-1 and Frame-2, can improve $S/B$ by about 15\%.
 More modest improvements can be obtained 
 in the laboratory frame.

\section{The $t\bar{t}b\bar{b}$ channel}
We now turn to the spin-correlated analysis for  
$t\bar{t}H (H\to b\bar{b})$. With respect to the  $H\to \gamma\gamma$
channel, here the  ($t\bar{t}b\bar{b}$) irreducible background receives contributions from many components that have different top spin correlations.
 This makes even more difficult to guess by general arguments how the 
correlated/uncorrelated background distributions may behave,
especially when kinematical cuts can  affect  predictions.
We stress that the following analysis is an idealized one where we assume to be able to distinguish the $b$ quarks coming from top (anti top) decays.

\begin{figure}[t]
\includegraphics[width=0.49\textwidth]{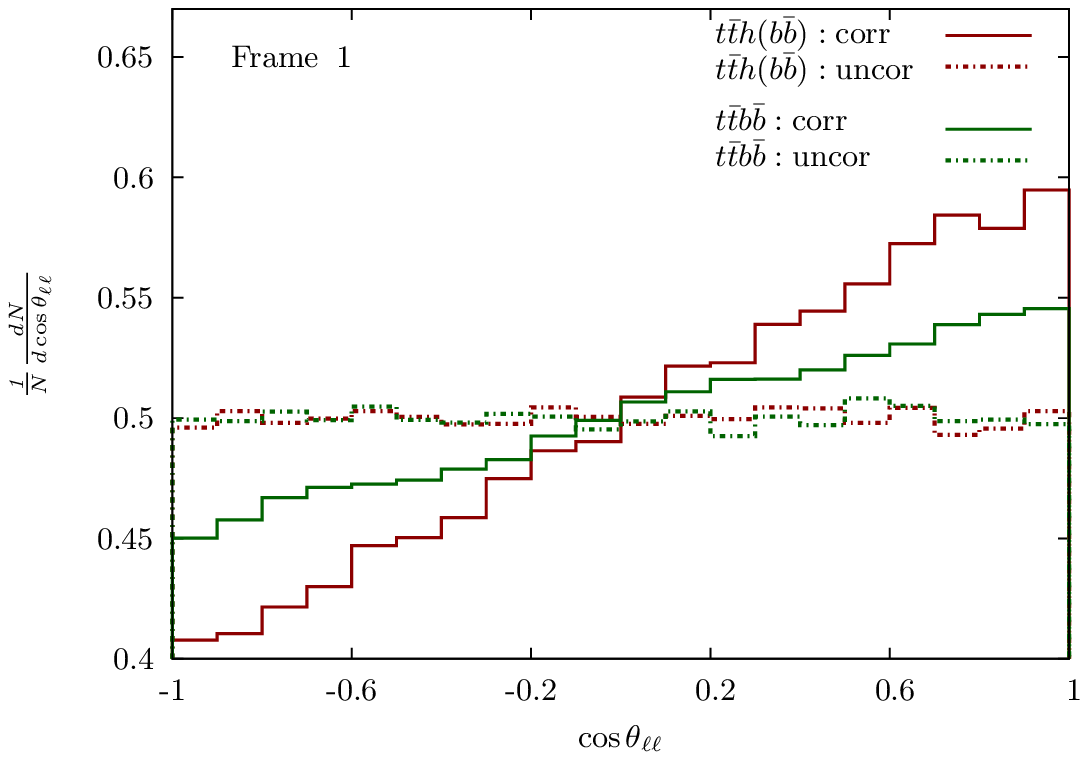}\hskip 9pt
\includegraphics[width=0.49\textwidth]{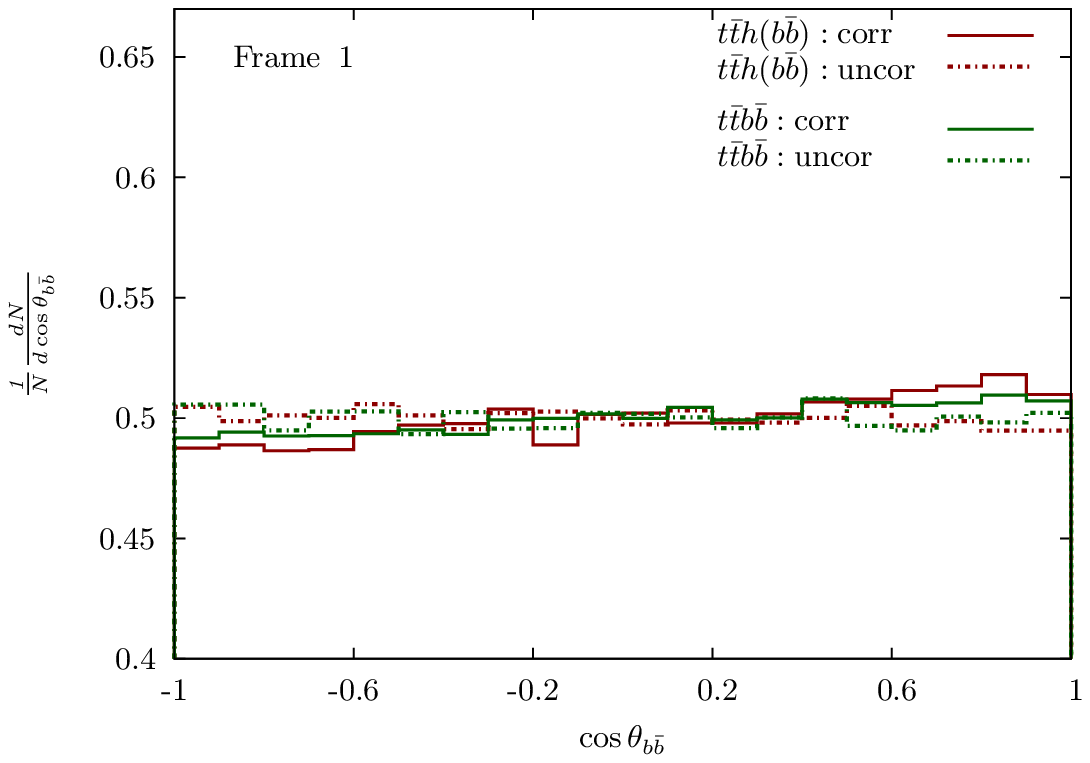}
\vskip 5pt
\includegraphics[width=0.49\textwidth]{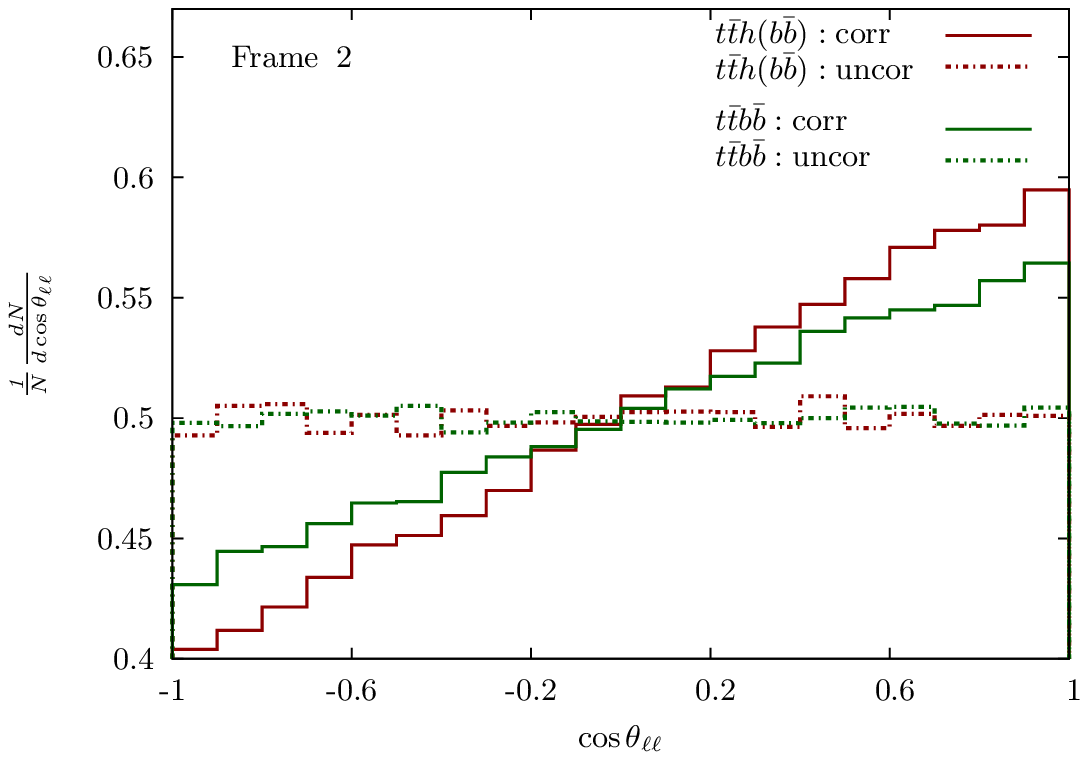}\hskip 9pt
\includegraphics[width=0.49\textwidth]{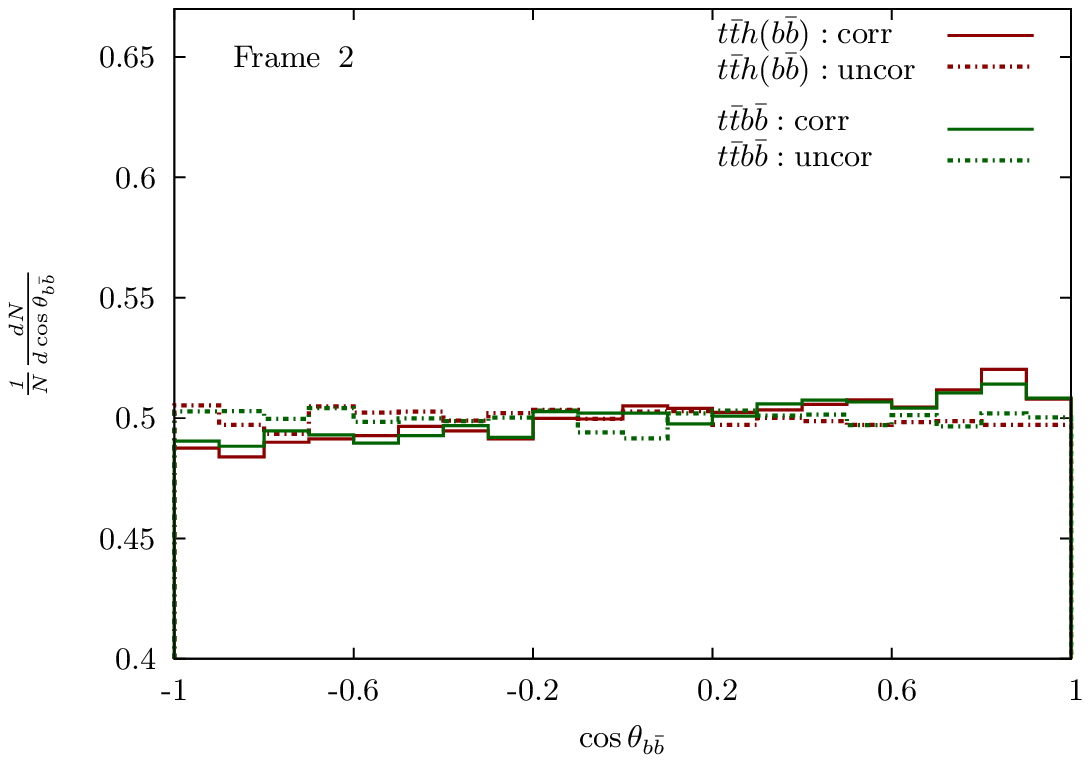}
\caption{ The 
$\ctll$ (left) and  $\ctbb$ (right)  
distributions for the signal $t\bar{t}H (H\to b \bar b)$ (red) and  $t\bar{t}b \bar b$ background (green), with (solid) and without (dashed) spin information, in  \text{Frame 1} (top) and \text{Frame 2} (bottom).
 The selection $p^{b,\bar{b}}_T > 20$ GeV, $|\eta_{b,\bar{b}}|<2.5$ and $\Delta R_{b\bar{b}} > 0.4$ has been imposed 
on the two  $b$-quarks not coming from $t$ decays, in addition to the invariant mass cut $m_{b\bar b}> 100$ GeV.}
\label{Fig7}
\end{figure}

In Fig.~\ref{Fig7}  we show the results for the 
$\ctll$ (left) 
 and $\ctbb$ (right) distributions 
for  signal and irreducible background, in Frame-1 (top) and Frame-2 (bottom).
In both frames spin effects are quite relevant in the $\ctll$ distributions. In particular, while the uncorrelated distributions are in general flat, 
both the  correlated signal and background  behaviors 
are  monotonically increasing as a function of $\ctll$, with a stronger slope in the signal case.
On the other hand, the effect of spin correlations is extremely mild for $\ctbb$ distributions in both Frame-1 and Frame-2. 
 \\
Then, in the $t\bar{t}H (H\to b\bar{b})$  case,  
retaining the full spin information  can still help in enhancing $S/B$, although  not as much as in the case of the $H\to \gamma \gamma$ channel.  In particular, in Frame-1, including spin correlations improves $S/B$ by 4\% using the $\ctll$ variable, while a very small gain is obtained by using $\ctbb$.

\begin{figure}[ht]
\includegraphics[width=0.49\textwidth]{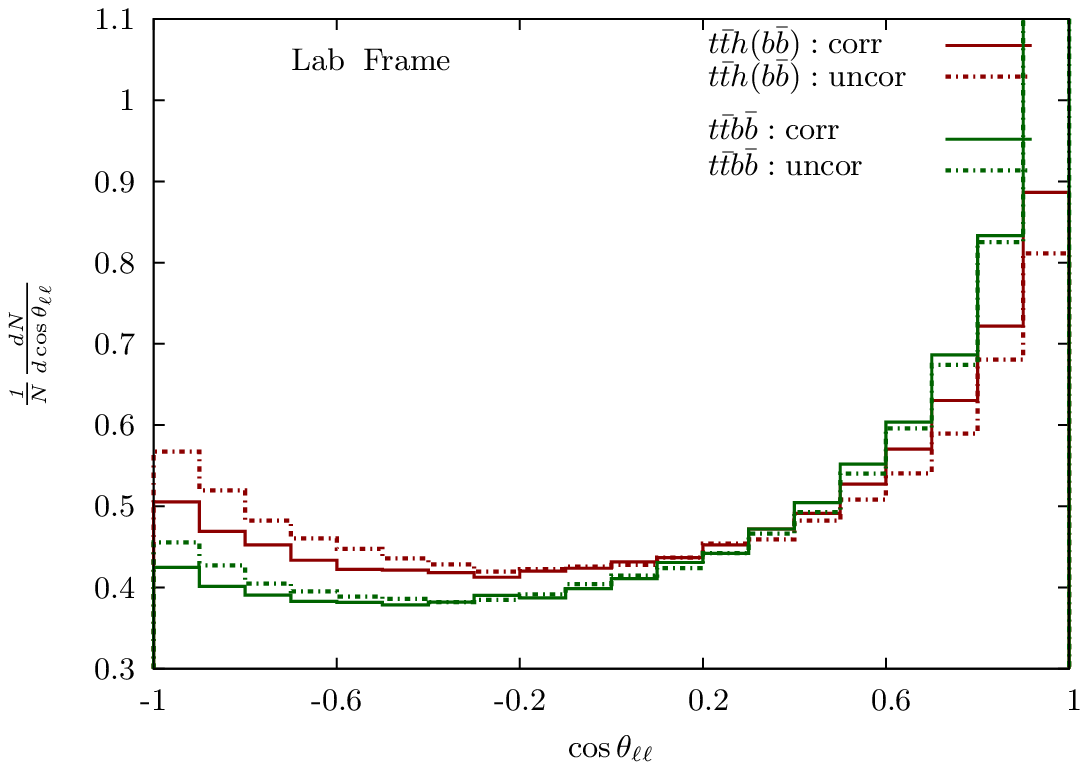}\hskip 9pt
\includegraphics[width=0.49\textwidth]{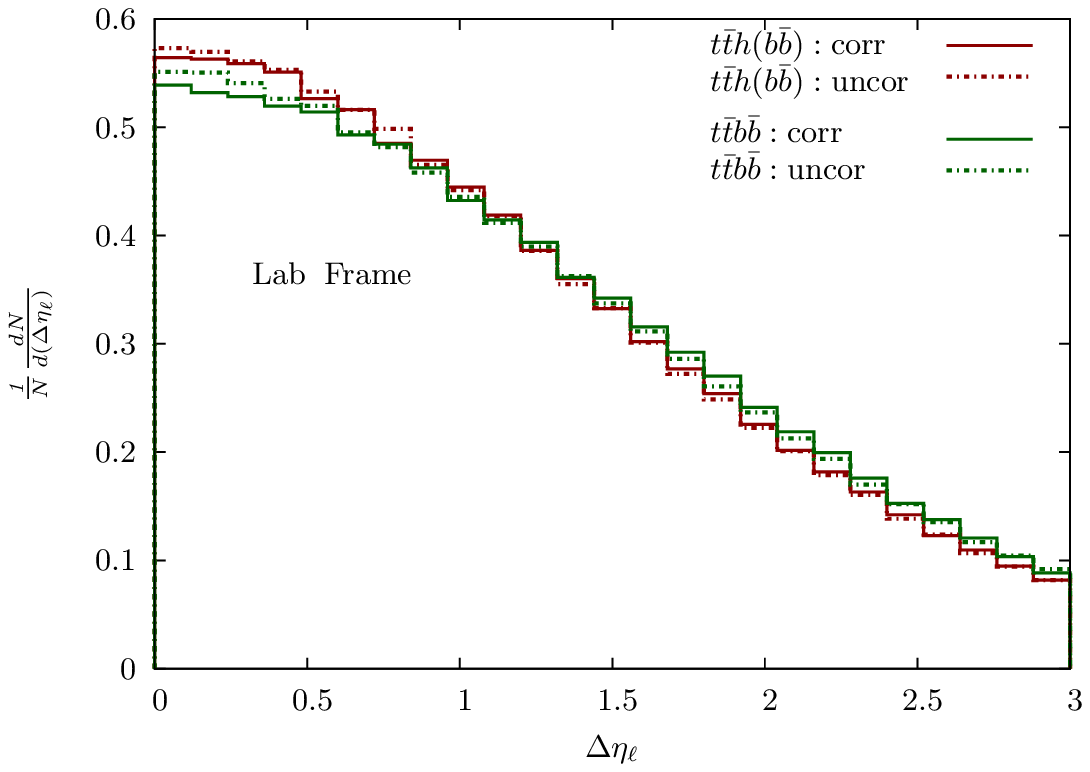}
\vskip 5pt
\includegraphics[width=0.49\textwidth]{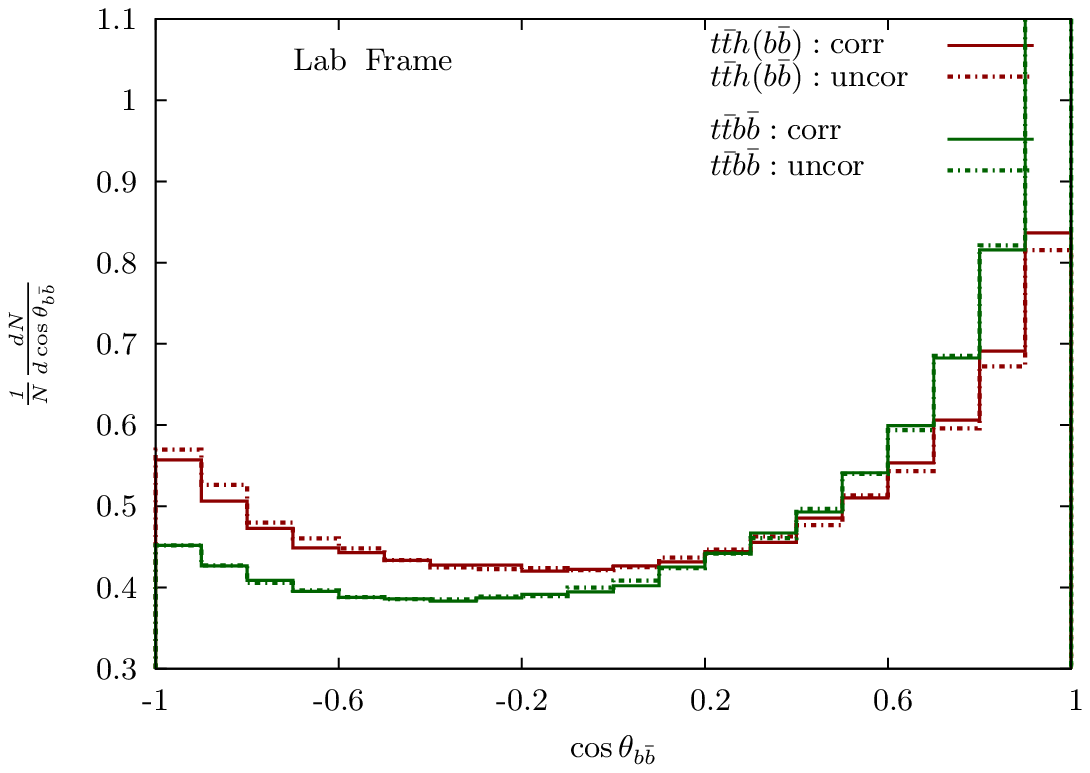}\hskip 9pt
\includegraphics[width=0.49\textwidth]{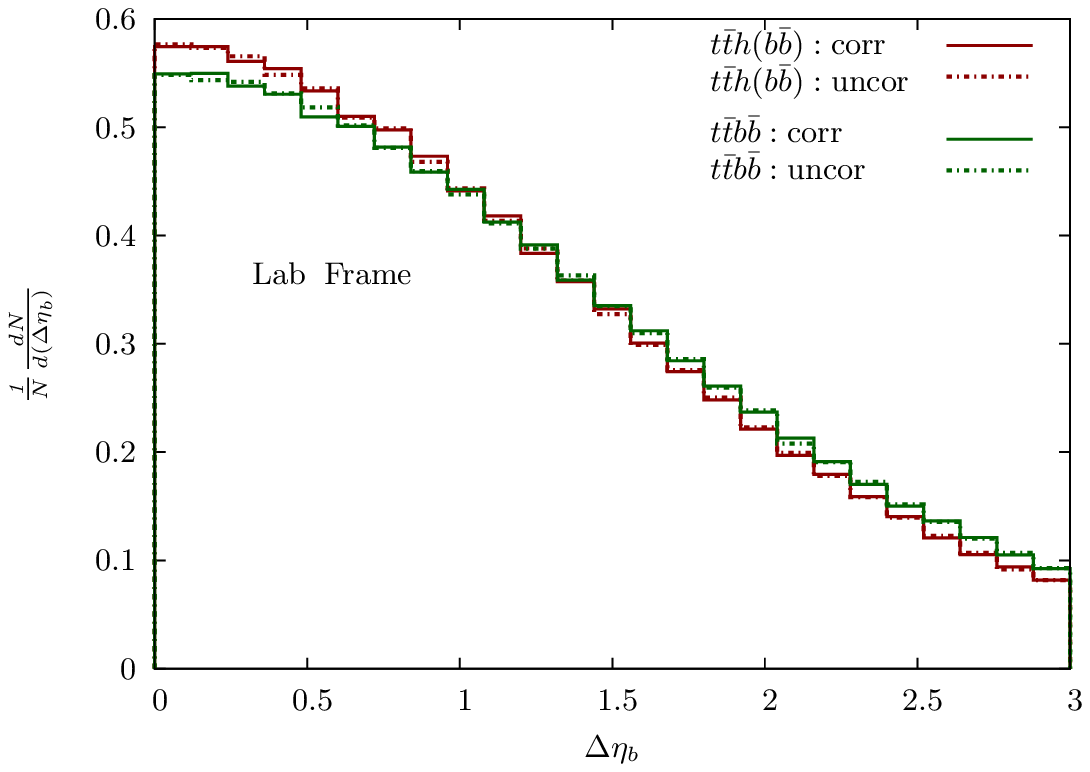}
\caption{ The 
$\ctll$ (top left), $\detall$ (top right),  $\ctbb$ (bottom left),  and $\detabb$ (bottom right) 
distributions for the signal $t\bar{t}H (H\to b\bar{b})$ (red) and  $t\bar{t}b\bar{b}$ background (green), with (solid) and without (dashed) spin information, in the \text{Lab frame}. Same cuts as in Fig.~7 have been imposed.}
\label{Fig8}
\end{figure}

For comparison we present in Fig.~\ref{Fig8} 
the corresponding results for the  $\ctll$ (top left), $\detall$ (top right),
$\ctbb$ (bottom left)  
and $\detabb$ (bottom right) distributions  in the laboratory frame. 
Spin effects are quite milder in this case, too, and in general do not improve much the signal-background separation.
Signal and background have similar distributions,  in both correlated and uncorrelated cases.
The advantage  
of employing a full spin-correlated analysis in the Lab frame looks quite modest.  
\begin{figure}[h]
\includegraphics[width=0.48\textwidth]{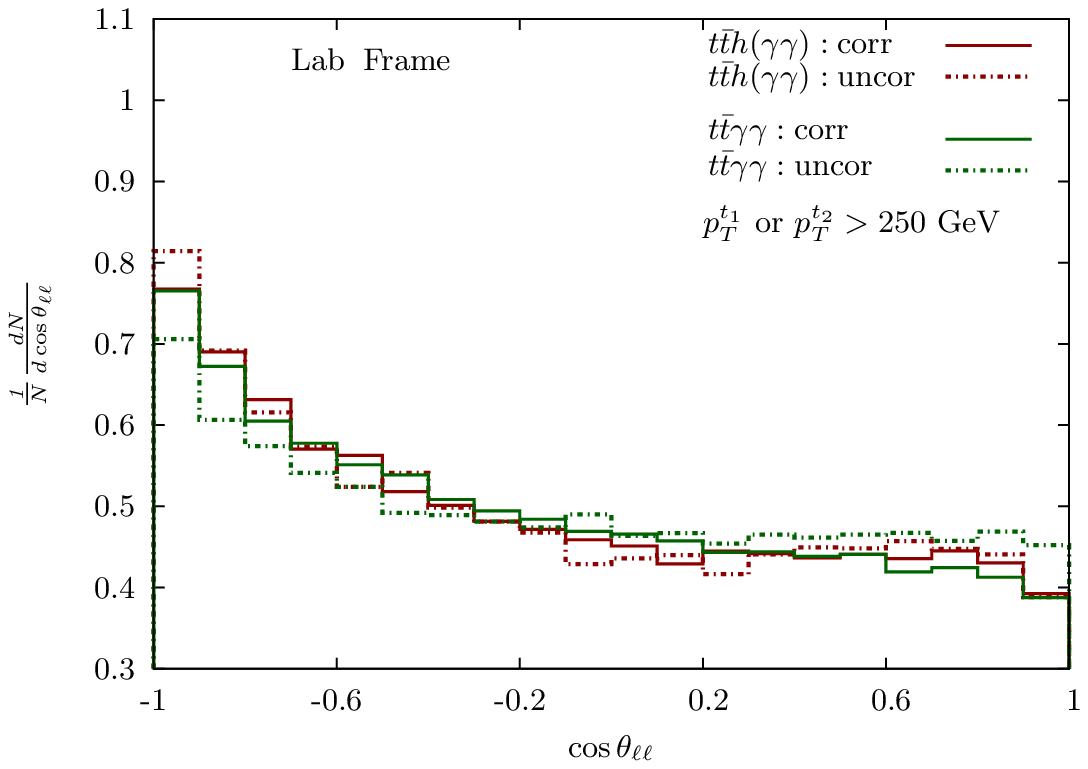}\hskip 9pt
\includegraphics[width=0.48\textwidth]{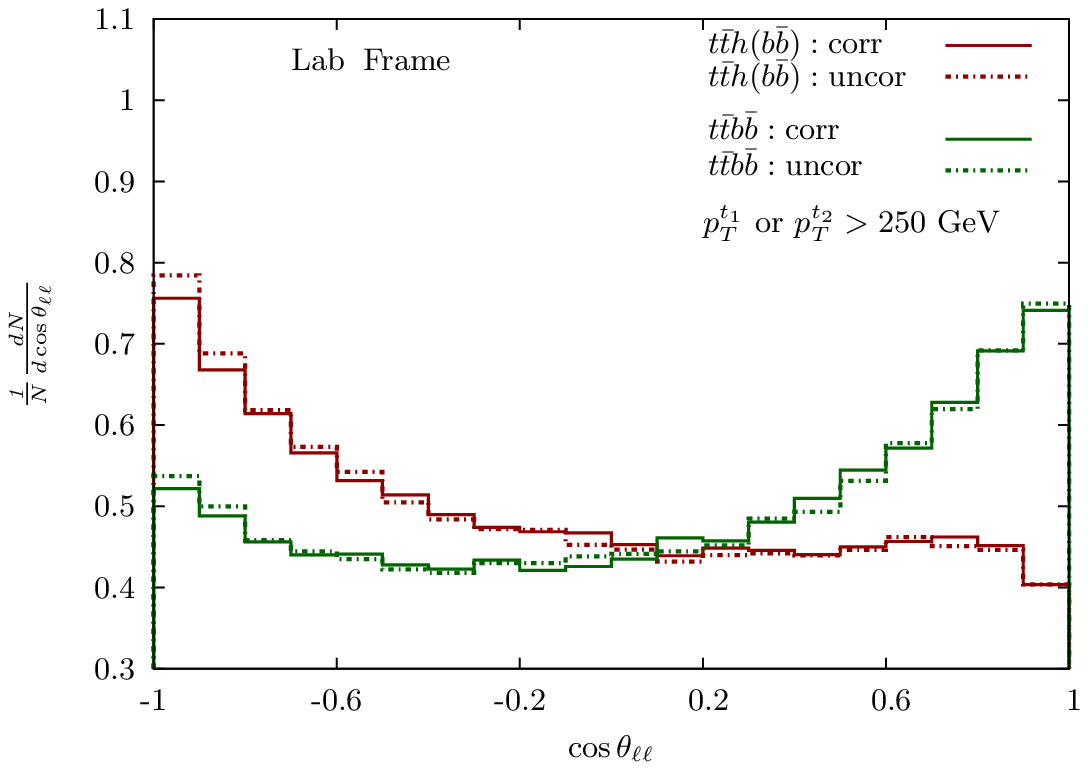}
\caption{The $\ctll$ distributions for the signals (red)  
$t\bar{t}H~ (H\to \gamma \gamma)$ (left) and 
$t\bar{t}H~(H\to b\bar{b})$ (right) with their corresponding backgrounds (green) $t\bar{t}\gamma \gamma$ and $t\bar{t}b\bar{b}$
in the Lab frame, after demanding one highly-boosted top by imposing that the highest  top  $p_T$ satisfies the cut $p_T> 250$ GeV. Same cuts as in Fig.~2 and 7 have been applied, respectively.}
\label{Fig9}
\end{figure}

Finally, in Fig.~\ref{Fig9} we explore  the boosted  
top regime, by imposing that the highest  top  $p_T$ satisfies the cut $p_T> 250$ GeV
 in the laboratory frame. 
In particular, we plot the $\ctll$ distributions for the signals  
$t\bar{t}H~ (H\to \gamma \gamma)$ (left) and 
$t\bar{t}H~(H\to b\bar{b})$ (right) with their corresponding backgrounds $t\bar{t}\gamma \gamma$ and $t\bar{t}b\bar{b}$,
 imposing the cuts $p_T > 20$ GeV,
$|\eta|<2.5$ and $\Delta R > 0.4$ on 
the final state photons or  $b$ quarks  (where  $b$'s are not from top decays) in addition to the invariant mass cut 
123 GeV$<m_{\gamma\gamma}<$ 129 GeV, and $m_{b\bar b}> 100$ GeV,
respectively.
The left and right plots  in Fig.~\ref{Fig9} should be compared with the top-left plots 
of Fig.~\ref{Fig4} and  Fig.~\ref{Fig8}, respectively.
Requiring a boosted top increases the lepton-pair angular separation in general.
Then, in the $H\to \gamma \gamma$, one obtains practically identical  uncorrelated curves, especially in the correlated case.  
 There is instead some   advantage for the $H\to b\bar{b}$ signal, where the separation 
 of the $\ctll$ distributions for signal and background increases both in the correlated and uncorrelated case. The spin-correlation effects are anyhow modest even in this case

Summing up, as for the  $H\to \gamma\gamma$ signal, we can see that
the most significant deviations between the correlated and 
uncorrelated analysis is observed in the 
$\ctll$ distributions, evaluated 
in { Frame-1} and { Frame-2}. On the other hand, for the $H\to b\bar{b}$ signal, 
the signal and background including spin correlations have similar (less distinctive) behaviors with respect to the $H\to \gamma\gamma$ case. Anyhow, 
 also for the $H\to b\bar{b}$ signal, 
an analysis taking into account spin correlations in a suitable frame could significantly help in enhancing the $S/B$ ratio with respect to the uncorrelated analysis.

\section{Summary and Outlook}

The top-quark polarization observables are quite powerful tools that can be used to enhance the sensitivity to the dynamics involved in the top-production processes. 
The main purpose of the present study was to investigate the 
advantages of taking into account the full $t \bar t$ spin-correlation effects in the measurement of the $t\bar{t}H$ process versus its irreducible backgrounds. 

We found that, for the two processes 
  $t\bar{t}H(H\to \gamma \gamma)$  and 
$t\bar{t}H(H\to b\bar{b})$, 
where irreducible backgrounds are bound to have a dominant role when increasing the LHC data set at 14 TeV, there are indeed angular variables defined in dedicated reference frames that could sizably increase the separation of signal and background, with a gain of up to 30\% in $S/B$ for 
 $t\bar{t}H(H\to \gamma \gamma)$ in particular phase-space regions.

Of course, our study suffers from a series of limitations that one will have to address in order to assess the actual potential of the suggested optimization strategy:
\begin{itemize}
\item[ ]
-- First, we just assumed a  simplified framework not including NLO QCD corrections and  parton-shower effects.
In \cite{Artoisenet:2012st}, one 
can see that NLO corrections tend to modify  the (uncorrelated) signal  in the same direction as the LO spin correlation effects. One should then also estimate the NLO corrections for the corresponding backgrounds, and confront them  with spin effects.

--  In the present study, we started to examine the effects of additional experimental kinematic selection cuts on spin correlations. We found that, in the $t\bar{t}H (H\to \gamma\gamma)$ channel,  realistic cuts do not upset the correlation behavior. On the other hand, a general depletion of the spin effects is observed in the background case. We expect a similar sensitivity to further cuts in the $t\bar{t}H (H\to b\bar{b})$ case.

-- We have assumed a 100\% top-system reconstruction efficiency, although we are considering the challenging dilepton final state containing two neutrinoÕs.  Our results could then be quite optimistic. In a more realistic experimental set-up,  the effects we found could be partly washed out by detection and resolution experimental effects, affecting  the reconstruction of the two $t$ and $\bar t$ 
rest frames.
  In \cite{Baumgart:2012ay},  these issues 
were discussed for the $t\bar t$ production. In the $t\bar th$ case, the lower production statistics will make the top reconstruction even harder.
 On the other hand, whenever one consider  spin-correlation distributions in the Lab frame (cf. Fig.~\ref{Fig4} and Fig.~\ref{Fig8}), the $t$ and $\bar t$ rest frame reconstruction is not needed, and the   spin-correlation results will not be deteriorated.
 
 -- In the $t\bar{t}H (H\to b\bar{b})$ case, we included only the irreducible $t\bar{t}b\bar{b}$ background, while by relaxing the $b$-tagging multiplicity  the reducible light-jet $t\bar{t}jj$  channel becomes  overwhelming.
We checked how the latter background reacts to the 
inclusion of $t\bar t$ spin correlations by using similar selection criteria than the ones in the $t\bar{t}b\bar{b}$ analysis. We found that, in Frame 1, the $\cos \theta_{\ell\ell}$ 
distribution for $t\bar{t}jj$ is similar to the $t\bar{t}b\bar{b}$ distribution, and approaches  the signal one. Spin correlations might  then be as much  effective as in the irreducible $t\bar{t}b\bar{b}$ channel for  separating signal from background.

-- In this study, we explored the  spin correlations in the 
dilepton $t\bar t$  channel, which is only a subdominant component of the $t\bar{t}H$ sample.
The possibility to include in spin-correlation studies the higher-rate lepton+jets channel   was studied for $t\bar t$ production in  \cite{Mahlon:2010gw,Baumgart:2012ay}.
Although the $W\to jj$ light-jet analyzing power is on average smaller, and the light-jet identification can be nontrivial (implying shower and hadronization distortions), for the semi-leptonic $t\bar t$ system  
the top  reconstruction turns out to be in general quite efficient. Similar techniques could be extended to the  $t\bar{t}H$ production.

--  The actual spin-correlation distributions of the $t\bar{t}H(H\to b\bar{b})$   signal will also be  affected by the $b$-jet combinatorial background.
A preliminary simulation that uses simplified assumption on top reconstruction
effects (in particular a 10 GeV mass resolution on the two top ($\ell \nu b$) systems, and a 15 GeV
resolution on the Higgs ($b\bar b$) system) results in :
a) an extra 10\% for the total event rates, and 
b) a few percents 
 distortion effect  on the signal $\cos \theta_{\ell\ell}$ distributions in Fig.~\ref{Fig7}, that tends to make the distribution slope flatter.
 \end{itemize}
%
%
%

We conclude that spin-correlation features in the $t\bar t H$ production are  quite 
promising for  enhancing the signal sensitivity over the irreducible background. Hence, they should 
definitely be 
studied in a more systematic way, and eventually be included in future analysis of the process  at higher integrated luminosities.

\vskip 0.4cm
{\bf Aknowledgments:} 
S.B. would like to thank Satya Mukhopadhay for helpful discussions. 
E.G. would like to thank the CERN PH-TH division for its kind hospitality during the preparation of this work. 
This work was supported by the ESF grant MTT60, by the recurrent financing SF0690030s09 project and by the European Union through the European 
Regional Development Fund.

\end{document}